\documentclass{pasa}
\usepackage{graphicx}
\usepackage[usenames,dvipsnames]{xcolor}
\usepackage{url}
\usepackage{aas_macros}
\usepackage{subcaption}
\usepackage{amsmath}
\usepackage{longtable}

\usepackage{hyperref}

\def\pasa{Publ. Astron. Soc. Australia}
\title[PPTA Data Release 2]{The Parkes Pulsar Timing Array Project: Second data release}
\author[M.~Kerr et al.]{
M.~Kerr$^{1}$\thanks{E-mail: matthew.kerr@nrl.navy.mil},
D.~J.~Reardon$^{2,3}$\thanks{E-mail: dreardon@swin.edu.au},
G.~Hobbs$^{4,3}$,
R.~M.~Shannon$^{2,3}$,
R.~N.~Manchester$^{4}$,
S.~Dai$^{4}$,
C.~J.~Russell$^{5}$,
 S-B.~Zhang$^{6,4,7,8}$,
 W.~van Straten$^{9}$,
S.~Os{\l}owski$^{2,3}$,
A.~Parthasarathy$^{2,3,4}$,
 R.~Spiewak$^{2,3}$,
M.~Bailes$^{2,3}$,
N.~D.~R.~Bhat$^{10}$,
 A.~D.~Cameron$^{4}$,
 W.~A.~Coles$^{11}$,
J.~Dempsey$^{12}$,
 X. Deng$^{4}$,
 B. Goncharov$^{13,3}$
 J.~F~Kaczmarek$^{14}$,
M.~J.~Keith$^{15}$,
 P.~D.~Lasky$^{13,3}$,
 M.~E.~Lower$^{2,3,4}$,
 B.~Preisig$^{14}$,
 J.~M.~Sarkissian$^{14}$,
L.~Toomey$^{4}$,
 H.~Wang$^{16}$,
 J.~Wang$^{17}$,
L.~Zhang$^{18,6,4}$,
X.~Zhu$^{13,3}$\\
{\em Affiliations are listed at the end of the paper.} 
}

\jid{PASA}
\doi{10.1017/pas.\the\year.xxx}
\jyear{\the\year}
\usepackage{aas_macros}

\begin{document}

\begin{frontmatter}
\maketitle
\begin{abstract}
We describe 14 years of public data from the Parkes Pulsar Timing Array (PPTA), an ongoing project that is producing precise measurements 
of pulse times of arrival from 26 millisecond pulsars using the 64-m Parkes radio telescope with a cadence of approximately three weeks in three observing bands. A comprehensive description
of the pulsar observing systems employed at the telescope since 2004 is provided, including the calibration methodology and an analysis of the stability of system components. 
We attempt to provide full accounting of the reduction from the raw measured Stokes parameters to pulse times of arrival
to aid third parties in reproducing our 
results. This conversion is encapsulated 
in a processing pipeline designed to track 
provenance. Our data products 
include pulse times of arrival for each of the pulsars along with an initial set of pulsar parameters and noise models.  The
calibrated pulse profiles and timing template profiles are also available.
These data represent almost 21,000\,hrs of recorded data
spanning over 14\,years.
After accounting for processes that induce time-correlated noise, 22 of the pulsars have weighted root-mean-square timing residuals
of $<1$\,$\mu$s in at least one radio band. The data should allow end users to quickly undertake
their own gravitational-wave analyses (for example) without having to understand the intricacies of pulsar polarisation calibration or attain a mastery of radio-frequency interference mitigation as is required when analysing raw data files. 
\end{abstract}

\end{frontmatter}

\section{Introduction}

A pulsar timing array (PTA) consists of observations of an ensemble of millisecond pulsars with precision measurements of pulse times of arrival (ToAs) over a long data span. Timing-array experiments have made rapid strides in recent years. They offer a rich array of science targets. Observations of individual pulsars can yield neutron star mass measurements \citep{2016ARA&A..54..401O}, offer insight into binary evolution and millisecond pulsar formation, place constraints on models for neutron-star physics, and provide stringent tests of general relativity and equivalence principles \citep[e.g.][]{Will14}.  Once well-understood processes have been modelled (e.g. astrometric and
binary), studies of the correlations between the timing residuals of the pulsars allow for the  realization of a stable Galactic-scale clock \citep{Hobbs12,Hobbs19}, the characterisation of solar system dynamics \citep{Champion10, caballero2018}, searches for nanohertz-frequency gravitational wave  (GW) sources,  including gravitational-wave  backgrounds \citep{Shannon15,Lentati15,Nanograv11GWB} and single GW sources \citep{Zhu14,Babak16,aggarwal2019}.  Pulsar timing arrays are sensitive to the gravitational waves produced by binary supermassive black holes and complement ground-based gravitational wave detectors that detect mergers involving stellar-mass black holes, neutron stars, or both \citep{2016PhRvL.116f1102A, 2017PhRvL.119p1101A}.  Details are provided in the introduction of \citet{Manchester13} (hereafter M+13) as well as in reviews provided by Hobbs \& Dai (2017)\nocite{hd17}, \cite{Tiburzi18}, and  \citet{btc+19}. 

The Parkes Pulsar Timing Array (PPTA) project commenced in 2004 and observes millisecond pulsars south of $\delta=+22^\circ$, which is near the Parkes northern declination limit. To place this effort in context, the PPTA is one of three major PTAs: additionally, the European Pulsar Timing Array 
combines data from the European large radio telescopes \citep[EPTA,][]{EPTA}, and North American Nanohertz Observatory for Gravitational Waves \citep[NANOGrav,][]{Nanograv11} uses  the Green Bank Telescope and Arecibo Telescope to time pulsars mostly north of $\delta>-40^\circ$. Historically, the Parkes 64\,m telescope has had almost
exclusive access to pulsars south of $\delta<-40^\circ$, giving it an important role in global efforts to detect gravitational waves. The
PPTA, NANOGrav and EPTA have combined to form the International Pulsar Timing Array (IPTA) \citep{verbiest2016,IPTAdr2} which has the detection of nanohertz gravitational waves as its primary goal.

The first major PPTA data set (Data Release 1, or DR1) was published by M+13 and contains data collected into 2011. 
That data set has been used in a variety of analyses, including producing limits on individual sources of ``continuous'' GWs \citep{Zhu15}, searching for ``gravitational wave memory'', the permanent imprint on spacetime of a black hole merger \citep{Wang15}, and constraining the stochastic GW background from the merger of cosmological supermassive black holes \citep{shannon13}.  The data set has also been used in many other studies, including to determine ties between celestial reference frames \citep{wch+17}, to measure pulsar rotation measures \citep{ymh+11}, and to study the interstellar medium \citep{you2017,kcs+13,coles15}.  

Since DR1, the PPTA team has continued to record and process further observations.  The initial data release was improved by \cite{Reardon16} who produced new noise models, exploiting the generalized least-squares method developed by \cite{Coles11}, and updated pulsar-timing model parameters. \cite{Shannon15} investigated  a longer data set, using a subset of the most precisely timed pulsars,  restricted to observations in a single frequency band, and provided stringent constraints on the amplitude of a GW background.  This data set was also combined with early ground-based limits on the stochastic background to constrain early-Universe sources of gravitational radiation \citep{lms+16}.
The data set was further updated and used by \cite{pzl+18}  to constrain a model of dark matter and a recent version of the PPTA data set has been published as part of the second data release for the International Pulsar Timing Array (IPTA) project  \citep{IPTAdr2}.

\begin{table*}
\begin{center}
\caption{ Fundamental parameters of the PPTA DR2 pulsars, including pulse period ($P$), dispersion measure (DM), and orbital period ($P_{\rm b}$). Pulse widths are derived from the mean pulse profile and are given for the 10\% and 50\% levels ($W_{10}$ and $W_{50}$ respectively) relative to the observed pulse peak. Flux densities ($S_f$ for centre frequency $f$) are represented by their median (med.), mean ($\mu$), and standard deviation estimate ($\sigma$, defined as half of the range between the 84th percentile flux and the 16th percentile flux), which were derived from the distributions of fluxes measured with the latest observing systems.}\label{tb:basicParams}
\begin{tabular}{lrrrrrrrr}
\hline
Pulsar name & $P$ & DM & $P_{\text{b}}$ & $W_{10}$ & $W_{50}$ & $S_{700}$ & $S_{1400}$ & $S_{3100}$ \\
J2000 & (ms) & (cm$^{-3}$pc) & (d) & (ms) & (ms) & (mJy) & (mJy) & (mJy) \\
 &  &  & &  &  & med., $\mu$, $\sigma$ & med., $\mu$, $\sigma$ & med., $\mu$, $\sigma$ \\
 \hline
 J0437$-$4715 & 5.76 & 2.6 & 5.74 & 1.01 & 0.14 & 300, 369, 232 & 158, 162, 37 & 32.2, 32.5, 2.6 \\  
 J0613$-$0200 & 3.06 & 38.8 & 1.20 & 0.93 & 0.47 & 6.7, 6.8, 1.6 & 2.2, 2.2, 0.4 & 0.4, 0.4, 0.1 \\
 J0711$-$6830 & 5.49 & 18.4 & --- & 2.56 & 1.90 & 7.2, 8.4, 5.0 & 2.1, 3.2, 2.2 & 0.5, 0.6, 0.3 \\ 
 J1017$-$7156 & 2.34 & 94.2 & 6.51 & 0.14 & 0.07 & 2.4, 2.4, 0.6 & 0.9, 0.9, 0.4 & 0.1, 0.2, 0.1 \\ 
 J1022+1001 & 16.45 & 10.3 & 7.81 & 1.97 & 0.97 & 6.2, 10.1, 8.3 & 3.1, 5.1, 3.6 & 1.1, 1.1, 0.3 \\
 \\
 J1024$-$0719 & 5.16 & 6.5 & --- & 1.50 & 0.51 & 3.3, 4.3, 3.1 & 1.6, 1.9, 1.1 & 0.4, 0.4, 0.1 \\ 
 J1045$-$4509 & 7.47 & 58.1 & 4.08 & 1.44 & 0.76 & 9.0, 9.0, 1.8 & 2.7, 2.8, 0.5 & 0.4, 0.4, 0.1 \\ 
 J1125$-$6014 & 2.63 & 53.0 & 8.75 & 1.51 & 0.14 & --- & 0.9, 1.0, 0.5 & 0.3, 0.4, 0.3 \\ 
 J1446$-$4701 & 2.19 & 55.8 & 0.28 & 0.28 & 0.08 & 1.7, 1.7, 0.5 & 0.4, 0.4, 0.1 & 0.2, 0.2, 0.1 \\ 
 J1545$-$4550 & 3.58 & 68.4 & 6.20 & 0.56 & 0.13 & 1.0, 1.0, 0.2 & 1.0, 1.0, 0.2 & 0.3, 0.3, 0.1 \\ 
 \\
 J1600$-$3053 & 3.60 & 52.3 & 14.35 &  0.41 & 0.09 & 2.6, 2.7, 0.5 & 2.5, 2.5, 0.4 & 0.8, 0.8, 0.2 \\ 
 J1603$-$7202 & 14.84 & 38.1 & 6.31 & 1.72 & 1.21 & 11.2, 12.1, 4.5 & 3.5, 3.9, 1.7 & 0.3, 0.3, 0.2 \\ 
 J1643$-$1224 & 4.62 & 62.4 & 147.02 & 0.93 & 0.32 & 13.3, 13.1, 1.9 & 4.7, 4.8, 0.8 & 1.1, 1.1, 0.2 \\ 
 J1713+0747 & 4.57 & 15.9 & 67.83 & 0.39 & 0.11 & 7.6, 9.5, 5.1 & 7.4, 9.9, 6.5 & 2.0, 2.5, 1.6 \\ 
 J1730$-$2304 & 8.12 & 9.6 & --- & 1.72 & 0.97  & 10.5, 11.0, 3.1 & 3.8, 3.9, 1.9 & 0.8, 1.2, 0.8 \\ 
 \\
 J1732$-$5049 & 5.31 & 56.8 & 5.26 & 1.63 & 0.29 &  6.3, 6.2, 1.8 & 1.8, 1.8, 0.3 & 0.3, 0.4, 0.1 \\ 
 J1744$-$1134 & 4.07 & 3.1 & --- & 0.25 & 0.14 & 6.7, 7.5, 4.6 & 2.5, 3.7, 3.0 & 0.5, 0.7, 0.4 \\ 
 J1824$-$2452A & 3.05 & 119.9 & --- & 1.61 & 0.98 & 10.9, 10.4, 2.7 & 2.4, 2.4, 0.5 & 0.3, 0.3, 0.1 \\ 
 J1832$-$0836 & 2.72 & 28.2 & --- & --- & 0.86 &  2.8, 2.8, 0.3 & 1.1, 1.1, 0.3 & 0.4, 0.4, 0.1 \\ 
 J1857+0943 & 5.36 & 13.3 & 12.33 & 3.02 & 0.52 & 10.7, 10.7, 2.8 & 4.1, 4.8, 2.0 & 0.7, 1.0, 0.6 \\  
 \\
 J1909$-$3744 & 2.95 & 10.4 & 1.53 & 0.09 & 0.04 & 4.0, 4.7, 2.3 & 1.7, 2.3, 1.8 & 0.5, 0.7, 0.5 \\ 
 J1939+2134 & 1.56 & 71.0 & --- & 0.86 & 0.79 & 59.9, 61.5, 17.4 & 12.6, 14.0, 4.7 & 1.4, 1.5, 0.7 \\ 
 J2124$-$3358 & 4.93 & 4.6 & --- & --- & 0.51 & 9.6, 15.0, 11.8 & 4.5, 5.0, 2.4 & 0.6, 0.6, 0.1 \\ 
 J2129$-$5721 & 3.73 & 31.9 & 6.63 & 0.62 & 0.26 & 4.7, 5.2, 2.4 & 0.8, 1.2, 1.0 & 0.3, 0.3, 0.1 \\ 
 J2145$-$0750 & 16.05 & 9.0 & 6.84 & 4.17 & 0.34 & 16.5, 24.1, 17.9 & 5.9, 9.8, 8.2 & 1.5, 1.6, 0.6 \\ 
 \\
 J2241$-$5236 & 2.19 & 11.4 & 0.15 & 0.12 & 0.07 & 4.4, 8.4, 7.0 & 1.8, 2.1, 1.3 & 0.3, 0.3, 0.1 \\ 
 \hline
 \end{tabular}
 \end{center}
 \end{table*}

The primary aim of this paper is to make available  ToAs for the current PPTA pulsars and to describe the processing pipeline and the intermediate data products produced by the pipeline.  We have made substantial efforts to produce the new data set in a way that preserves the methodology and reasoning behind its processing. The paper gives an overview of the way we have recorded ancillary information and provides examples of the systematic errors we have identified in the data and their impact. Our results demonstrate the importance of detailed calibration and mitigation of radio-frequency interference (RFI).  Subsequent analyses  are already planned that will be based on the second Data Release (DR2) described here.  That future work will include studies of the noise present in the residuals, searches for gravitational waves and an analysis of the individual pulsars in the data set.

In \S\ref{sec:dataAcq} we describe the pulsars, the observing strategies and the receiver and signal-processor systems.   In \S\ref{sec:dataRed} we discuss our pipeline processing and calibration strategies. The resulting data set is presented and described in \S\ref{sec:dataSet}. We compare our data set with previous PPTA data releases in \S\ref{sec:discussion} as well as highlighting various features of our new release.  In the appendices, we present and describe example data files obtained from the processing pipeline and available in the data release.  Our data release is publicly available from \url{https://doi.org/10.25919/5db90a8bdeb59}.

The data products have been designed to be used with the {\sc Psrchive} \citep{Hotan04} and {\sc tempo2} \citep{hem06,ehm06} software packages. For the results presented here the ToAs were analysed using the Jet Propulsion Laboratory
solar-system ephemeris DE436\footnote{Available from \url{ftp://ssd.jpl.nasa.gov/pub/eph/planets/ascii/de436/}. See https://naif.jpl.nasa.gov/pub/naif/JUNO/kernels/spk/de436s.bsp.lbl for a brief description of DE436.} and the TT(BIPM18) reference timescale  published by the Bureau International des Poids et Mesures (BIPM).\footnote{Available from \url{ftp://ftp2.bipm.org/pub/tai/ttbipm/TTBIPM.2018}.}

\section{The PPTA Observation System and Pulsar Sample}\label{sec:dataAcq}

The observations described here were all obtained with the 64-m Parkes radio telescope in New South Wales, Australia.  PPTA observations commenced on 2004 February 6 and continue through to the present. Here, we include data from the onset of the project 2004 February 6 (MJD~53041) through to 2018 April 25 (MJD~58233), a span of 14.2\,yr, except for PSR~J0437$-$4715, where we provide early-science data from 2003 April 12 (MJD~52741), giving a span of $\sim 15$\,yr.

We have recently made a major upgrade of the observing system at the Parkes telescope to enable ultra-wide bandwidth observations, including a new ``UWL'' receiver, high-speed digitiser systems and a new signal-processor system based on graphics processor units (GPUs) \cite[][]{Hobbs19uwl}. The PPTA is now transitioning to these newer systems and will cease observations using the previous receivers and signal-processor systems once a sufficient overlap is obtained.  The data set that we describe here therefore includes all data prior to the installation and commissioning of the new receiver system during 2018.

Until the advent of the wide-band receiver system, we observed an ensemble of $\sim$24 millisecond pulsars typically every three weeks in three radio bands (10\,cm, 20\,cm, and 40/50\,cm). 
M+13 described the receiver and signal-processor instruments in detail. Since that time we have continued to use the 13-beam multibeam receiver \citep{swb+96} and the H-OH receiver in the 20\,cm observing band as well as the 10/40\,cm dual-band receiver \citep{gzg+01}.  Observations included here, but not in M+13, were obtained using the mark-3 and mark-4 versions of the Parkes Digital Filterbank Systems (PDFB3 and PDFB4, respectively) and the CASPER-Parkes-Swinburne recorder (CASPSR).  The last PDFB3 observation occurred in April 2014 when it suffered a hardware failure. Since then, until the advent of the UWL system, only PDFB4 and CASPSR were used.

Timing-array experiments rely on long-term timing observations of millisecond pulsars in which the various noise processes can be accurately modelled \citep{Coles11,VanHaasteren13,Lentati14}. For instance, dispersion measure (DM) variations can dominate timing residuals if multi-band observations are not available (see, e.g., \citealt{kcs+13}). This implies that, to be useful for the PPTA experiment, a pulsar must be sufficiently bright to obtain precise ToAs in at least two of the three observing bands.  It may be possible to correct for dispersion measure variations within individual bands of sufficient fractional bandwidth \cite[][]{Verbiest18}, which may have advantages to multi-band observations \citep{Pennucci14,Cordes16}.

A summary of the pulsars in this data release is given in Table~\ref{tb:basicParams}. This table contains the pulse periods, DMs, orbital periods, mean pulse widths for the 20\,cm band profile at 10\% and 50\% of the profile peak, and mean flux densities in the 40\,cm, 20\,cm and 10\,cm observing bands averaged over the available data span. Because of interstellar scintillation, observed flux densities vary from day to day, often by large factors. The quoted values are long-term averages, and we estimate that their systematic error is less than a few percent. 

During observing sessions three types of observations are conducted.
 The primary data product is pulsar fold-mode data. 
Immediately prior to each of these we observe a  pulsed noise diode, which is used for the calibration (see \S\ref{sec:calib}). This observation is slightly offset from the pulsar position to prevent the pulsar signal from contributing excess noise (which is particularly important for PSR~J0437$-$4715).  
For some of the pulsars with the most accurate arrival times, we also record the noise source after the observation, which enables interpolation of the calibration solution across the observation to correct for drifts. 
During most observing sessions we also observe the radio galaxy Hydra~A  (PKS 0915$-$11) as a primary flux density calibrator. The ``raw'' (uncalibrated) data files are available from the Parkes data archive (available from data.csiro.au; see \citealt{Hobbs11})\footnote{All observations included in DR2 are now outside the 18-month embargo period.}. Each observation was taken as part of an observing project that had a unique identifier. By far the majority of the observations were obtained as part of the formal PPTA project (P456), but a significant number were obtained as part two other projects.  P140 was an observing project to carry out high precision timing of millisecond pulsars in the 20-cm observing band and preceded the PPTA commencement.
P895 was started in order to carry out high cadence observations of PSR~J1909$-$3744 in the 10\,cm observing band. Some data were taken as part of project P865, which commensally searched for fast radio bursts while monitoring PPTA pulsars (see \citealt{oslowski2019}).  The data collection identifies from which project each observation belongs. We use the \textsc{Psrfits} data format \citep{Hotan04} which stores pulsar data along with observational metadata in binary Flexible Image Transport System (FITS) tables.

The observing strategy (see Figure 1)  evolved as new pulsars were added to the array and as the RFI environment changed. The fraction of the band that is clear is evolving with time, as terrestrial use of the band increases (e.g., mobile handsets) and more satellites are launched that use the 20cm band for transmission (for example GNSS services).
For the data presented here the primary strategy was to observe a pulsar in a given band for $\sim$5\,min. At this point the pulsar profile was inspected by eye.  If the pulsar was in a low scintillation state, or if the RFI was much stronger than typical, then the observation would be stopped and a new pulsar (or a new observing band) chosen. The initial pulsar would be re-observed after an interval greater than the diffractive time-scale (see, e.g., \citealt{you2017}) had elapsed.  

\begin{figure*}    
\centering
\includegraphics[width=14cm]{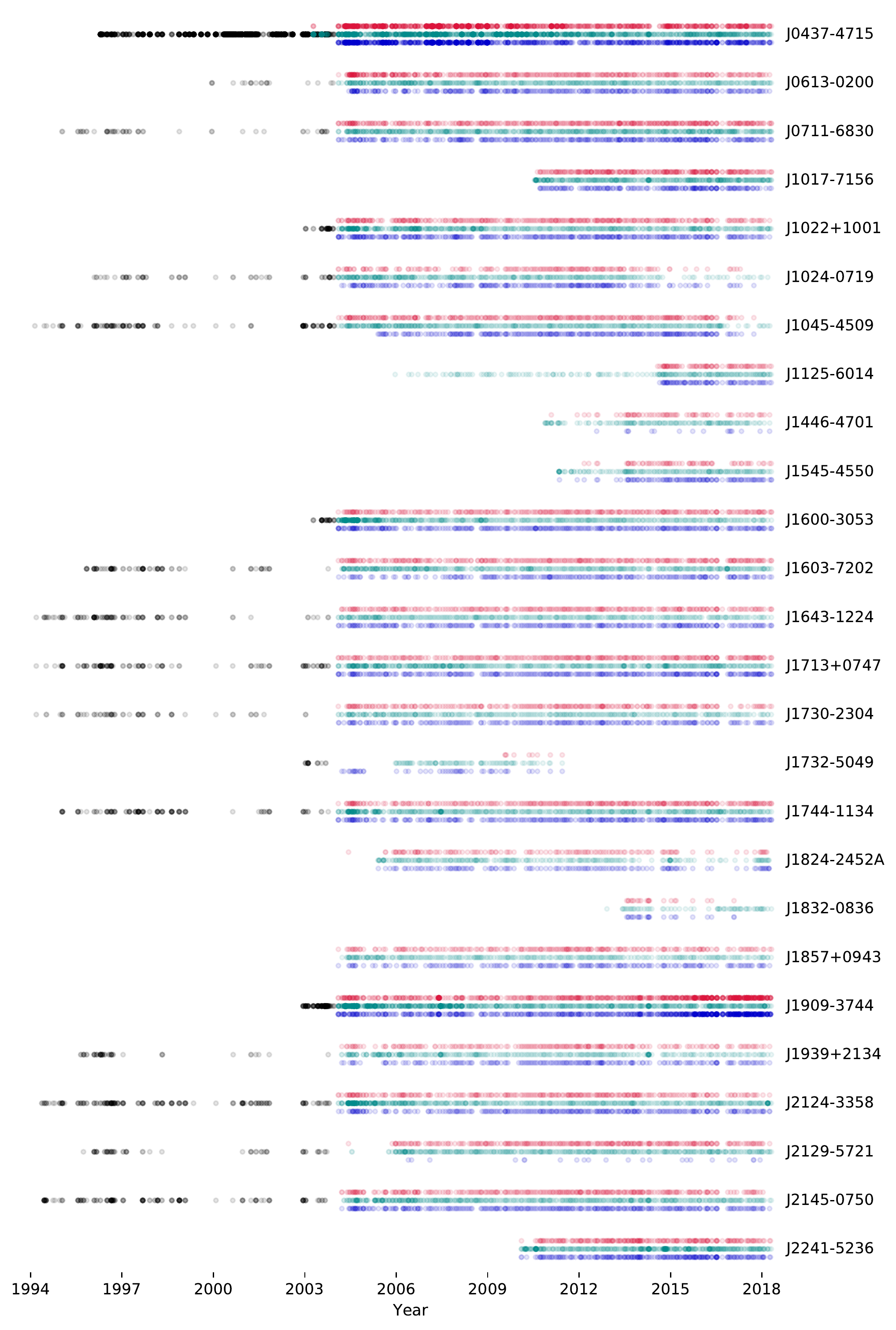}
\caption{The PPTA observing cadence. The three observing bands (10\,cm, 20\,cm and 40/50\,cm) are shown in blue, cyan, and red respectively, while the legacy 20\,cm observations are shown in black.}
\label{fg:cadence_dr2e}
\end{figure*}

Each pulsar is typically observed every two to three weeks.  In Figure~\ref{fg:cadence_dr2e} we show the observing cadence in the three observing bands.  The first PPTA data release included an extended version (known as ``DR1e'') with pre-2004 legacy 20\,cm data included, based on a data set published in \cite{Verbiest2009}.  We have not re-processed these earlier observations for DR2, but if the longer data spans are required then these observations can be obtained from the M+13 data collection. These earlier observations are not used for any of the subsequent work described in this paper, but their cadence is shown in Figure 1.

As new pulsars are added into the PPTA sample, the existing sample is inspected to see whether any pulsars should be removed. PSR~J1732$-$5049 was regularly observed from  2005 December until 2011 May, but the very wide pulse profile leads to relatively imprecise ToA measurements with typical uncertainties of 1 to 3\,$\mu$s.  The pulsar was subsequently removed from the array, but may be of interest for timing programmes on future high-gain southern hemisphere telescopes, such as MeerKAT \citep{bbb18} and the Square Kilometre Array (SKA; \citealt{Janssen15,keane18}).

PSR~J1832$-$0836 has been observed with modest cadence since June 2013.  It is relatively faint at both 10\, and 40\,cm, making DM variation correction difficult.  However, root-mean-square (rms) residuals below 1\,$\mu$s can still be achieved, owing to a profile with three narrow components,  and consequently, the PPTA observes this pulsar, but, until the advent of the UWL receiver, only in the 20-cm observing band.
These data can be combined with multi-band IPTA data, observed as part of the NANOGrav project \cite[][]{Nanograv11}.

\begin{figure*}
\includegraphics[width=4.1cm,angle=-90]{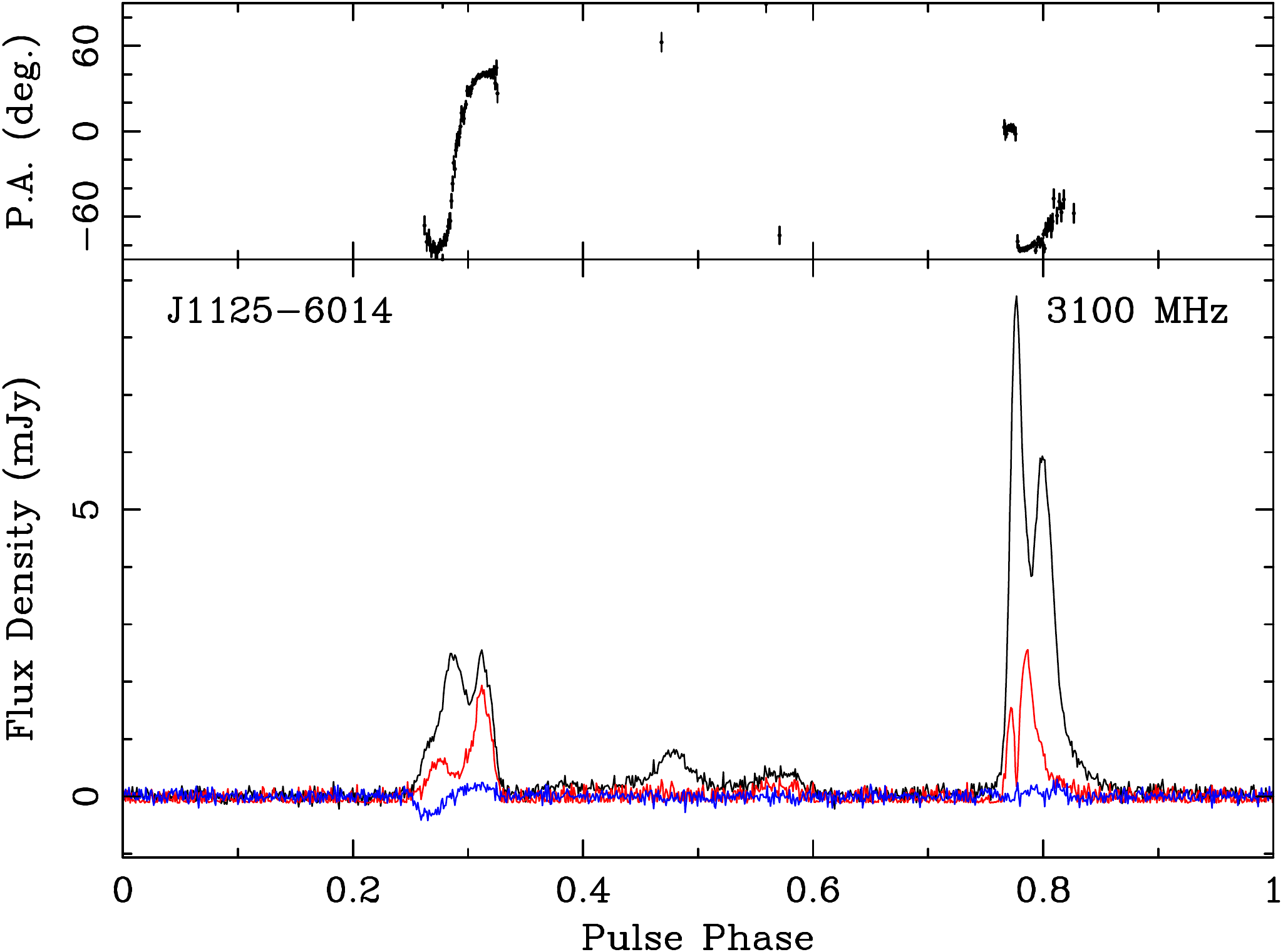}
\includegraphics[width=4.1cm,angle=-90]{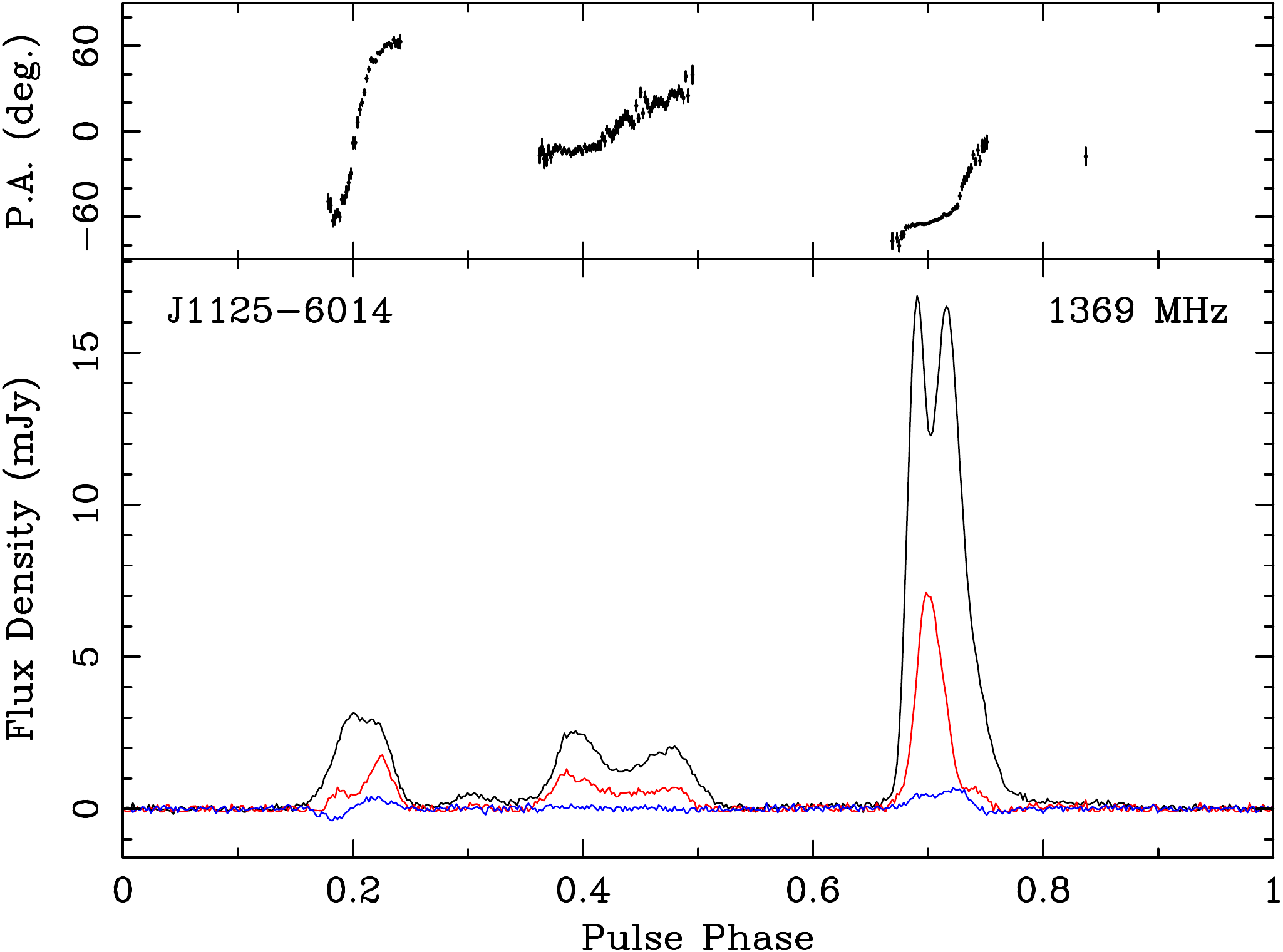}
\includegraphics[width=4.1cm,angle=-90]{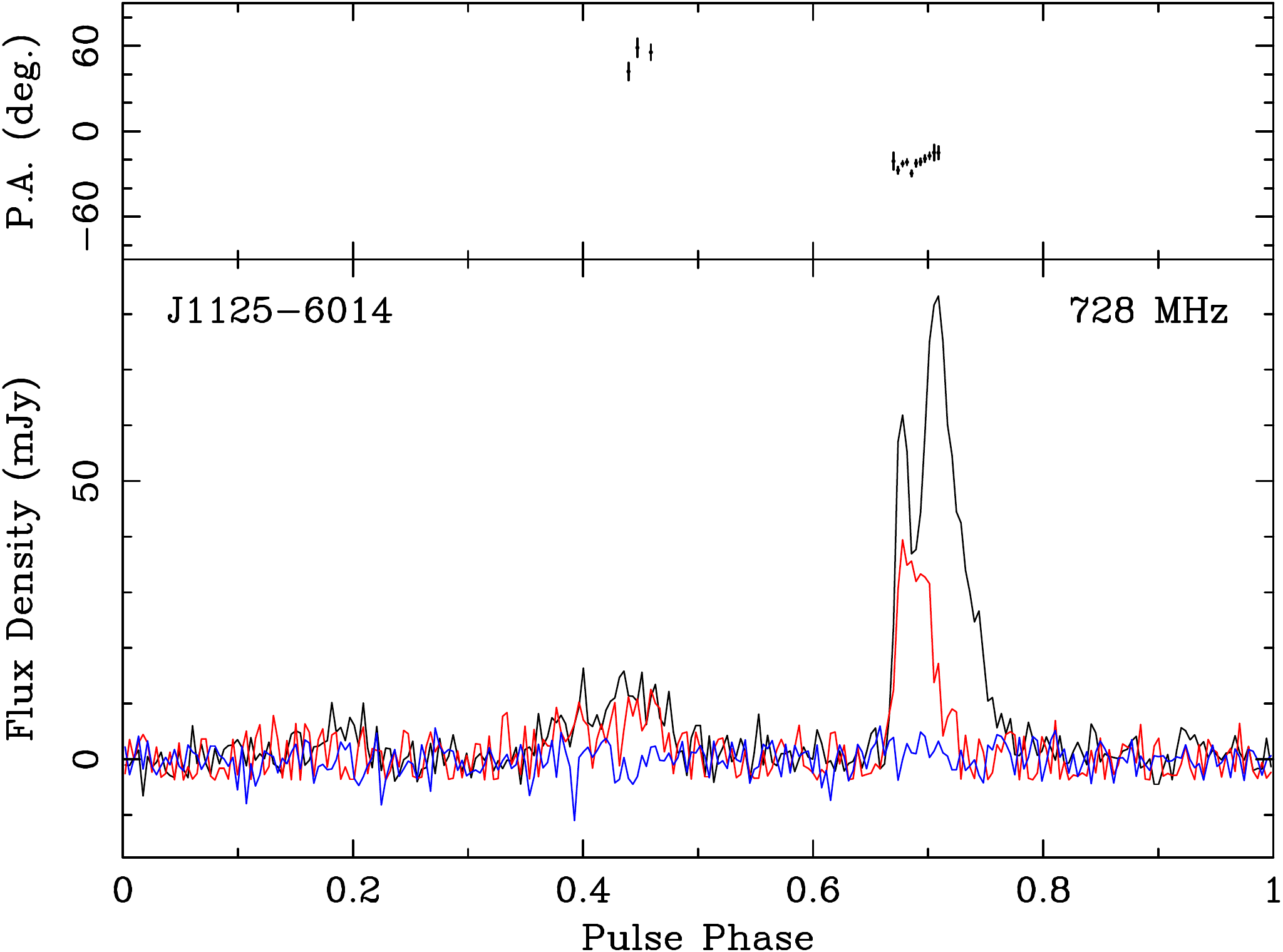}\\
\caption{Pulse profiles in the 10\,cm (left), 20\,cm (central) and 40/50\,cm (right) observing bands for PSR~J1125$-$6014. 
For each observing band we show the total intensity profile in black, the linear polarisation in red and the circular polarisation in blue.  The position angle of the linear polarisation is shown in the top sub-panel.}\label{fg:profiles}
\end{figure*}

Although it is an interesting source for the study of the solar wind (see, e.g., \citealt{ych+12}) and globular cluster dynamics, PSR~J1824$-$2452A boasts some of the largest timing and DM variation noise levels of any millisecond pulsar and is observed only with low priority.
Following the first release of PPTA data, five new pulsars have been added to the array as itemised below. Average pulse profiles for the new pulsars were analysed and published in \citet{Dai15}, with the exception of the profile of PSR~J1125$-$6014, shown in Figure \ref{fg:profiles}.

\begin{itemize}
\item PSR~J1017$-$7156 has a narrow pulse width allowing for precise ToAs to be determined (particularly in the 20\,cm observing band). Its discovery was presented in \citet{Keith12} in the HTRU Parkes pulsar survey and has been observed since July 2010.
\item PSR~J1125$-$6014 was found in the Parkes Mulitbeam Pulsar
Survey \citep{Lorimer06} and timed at 20\,cm only in a dedicated follow-up program (P501) and a general-purpose binary pulsar timing campaign (P789) from December 2005.  Multi-band PPTA observations commenced in July 2014.  
\item PSR~J1446$-$4701, discovered by \cite{Keith12}, is a $\gamma$-ray pulsar that has been observed since November 2011. It is a black-widow system with a 6.7\,hr orbit. 
\item PSR~J1545$-$4550 \citep{Burgay13} has been recorded with 20\,cm coverage since May 2011 and multi-wavelength PPTA coverage from February 2012. 
\item PSR~J2241$-$5236  was discovered in a targeted search of Fermi $\gamma$-ray sources by \citet{Keith11}. Early multi-wavelength observations in a variety of programs commenced February 2010, and regular PPTA observations commenced in August 2010. PSR~J2241$-$5236 could be classified as a black-widow system (although it shows no evidence of eclipse) with a 3.5-hr orbital period and an extremely light ($\sim 0.011$M$_\odot$) companion.  As discussed below, the pulsar shows orbit-induced timing variations.
\end{itemize}

\section{The Data Processing Pipeline}\label{sec:dataRed}

One of the primary aims of this data release is to ensure a transparent processing of the raw data into the final ToA products.  
We start from an archive of the raw data and provide a description of all steps undertaken in the analysis and in producing the reduced data.  We do this through the use of \textit{psrsh}, a domain-specific language for the \textsc{psrchive} package\footnote{\url{http://psrchive.sourceforge.net/manuals/psrsh}}.  \textit{psrsh} is a simple, human-readable set of commands that describe a data reduction pipeline.   The goal of our processing pipeline is thus to create optimal \textit{psrsh} scripts.  An example of such a script (and how it is used) is given in Appendix~\ref{sec:psrsh}.

The pipeline code is implemented in Python and, together with ancillary text files, is available as open source
software\footnote{\url{http://bitbucket.org/kerrm/dr2}}. In the following sub-sections we describe the principal components of the data-processing pipeline.   

\subsection{RFI mitigation}\label{sec:rfi}

The RFI environment at Parkes is surprisingly good, given that many licensed transmitters exist within the bands used for radio astronomy. However, 
RFI is present in nearly all PPTA data.  In this subsection, we give a brief overview of the interference environment and the mitigation steps that we have taken.

In mid-2009, as described in M+13, concern over digital television transmissions resulted in the 64-MHz 50\,cm band,  previously centred at 685\,MHz, being moved to 732\, MHz and hence renamed as the 40\,cm band.  After this move, the band was relatively clean apart from various relatively narrow-band transmissions which were excised using the  \textsc{psrchive} median bandpass filter implemented in the routine \textit{paz}.  For a number of years the band became cleaner as analogue television broadcasts were phased out.  
 
 In 2015, April, a 4G mobile phone transmitter in the band 758--768\,MHz began operation in Alectown, NSW, about 10\,km north of the telescope. These signals were sufficiently bright to introduce substantial system non-linearities.  To mitigate this, in 2015, May, we moved the centre frequency to 724\,MHz, despite this resulting in a somewhat sub-optimal bandpass. There is also occasional in-band RFI associated with 4G mobile phone handsets in the band 703--733\,MHz.  We excise affected sub-integrations that survive the spectral kurtosis and median bandpass filtering. Occasionally, most of the band is zero-weighted, and we discard such observations completely.

Within the nominal 256$-$350 MHz bandwidths of the 20\,cm systems, the band is relatively free from RFI during most observations.  There is a narrow-band microwave link operating around 1450\,MHz (1440\,MHz prior to MJD$\sim$55000) which we excise with the median bandpass filter. The most destructive source of  RFI is the episodic occurrence of aircraft radar, which can persist for several hours and, when strong, produces non-linearities in the low-noise amplifiers.   During the presence of the radar, observers typically switch to the dual band 10cm/40cm system.  
However, we discard observations that are badly affected by the interference.
More recently, and with higher cadence, RFI from satellites, particularly those of the {\it Beidou},  {\it COMPASS}, and {\it Galileo} global-positioning system constellations have affected our observations.  When these satellites pass through far side-lobes of the telescope beam, they contaminate their $\sim$\,60\,MHz bandwidth, but when they pass near the main beam, they result in nonlinearities. 
We also discard these latter observations. 

Finally, the 10-cm system has been largely free from RFI and an automatic median bandpass filter suffices to remove most instances.  Since mid-2015, several 20\,MHz-wide channels associated with mobile phone and data transmission licenses have become active.  The transmission strength has varied with date and with telescope position, and we have tabulated the affected observations to remove the appropriate channel range. 

Pipeline deletion or flagging of RFI is under control of ``rule sets'' which specify the frequencies and/or times affected by RFI for each observation and the mitigation method. For illustration, part of an RFI mitigation rule set is shown in Appendix~\ref{sec:ruleSet}.

\subsection{Calibration Methods}
\label{sec:calib}

To calibrate the system, we make use of a noise diode cycled at 11.123\,Hz and injected into the signal path for each polarisation, normally into the feed horn at 45$^\circ$ to each signal probe. This is used to calibrate the differential gain and phase of the two signal paths and, with observations of the flux-density calibrator (Hydra A), the equivalent flux density of both the calibration signal and the system.  See M+13 for a full description of the calibration system and the operational procedures for polarisation and flux density calibration. 

\begin{figure*}
\begin{subfigure}[t]{0.49\textwidth}
\includegraphics[angle=0,width=\textwidth]{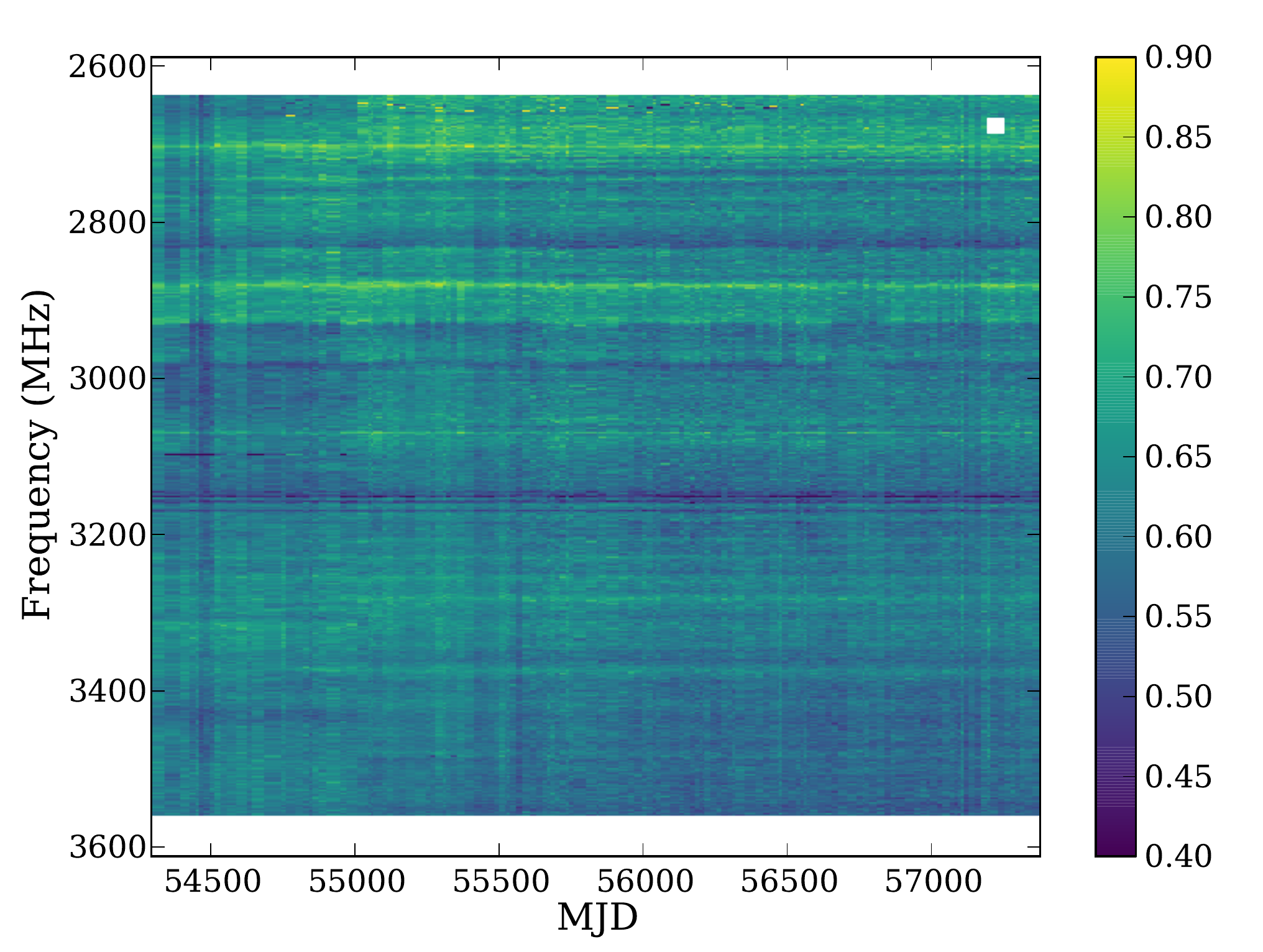}
\end{subfigure}
\begin{subfigure}[t]{0.49\textwidth}
\includegraphics[angle=0,width=\textwidth]{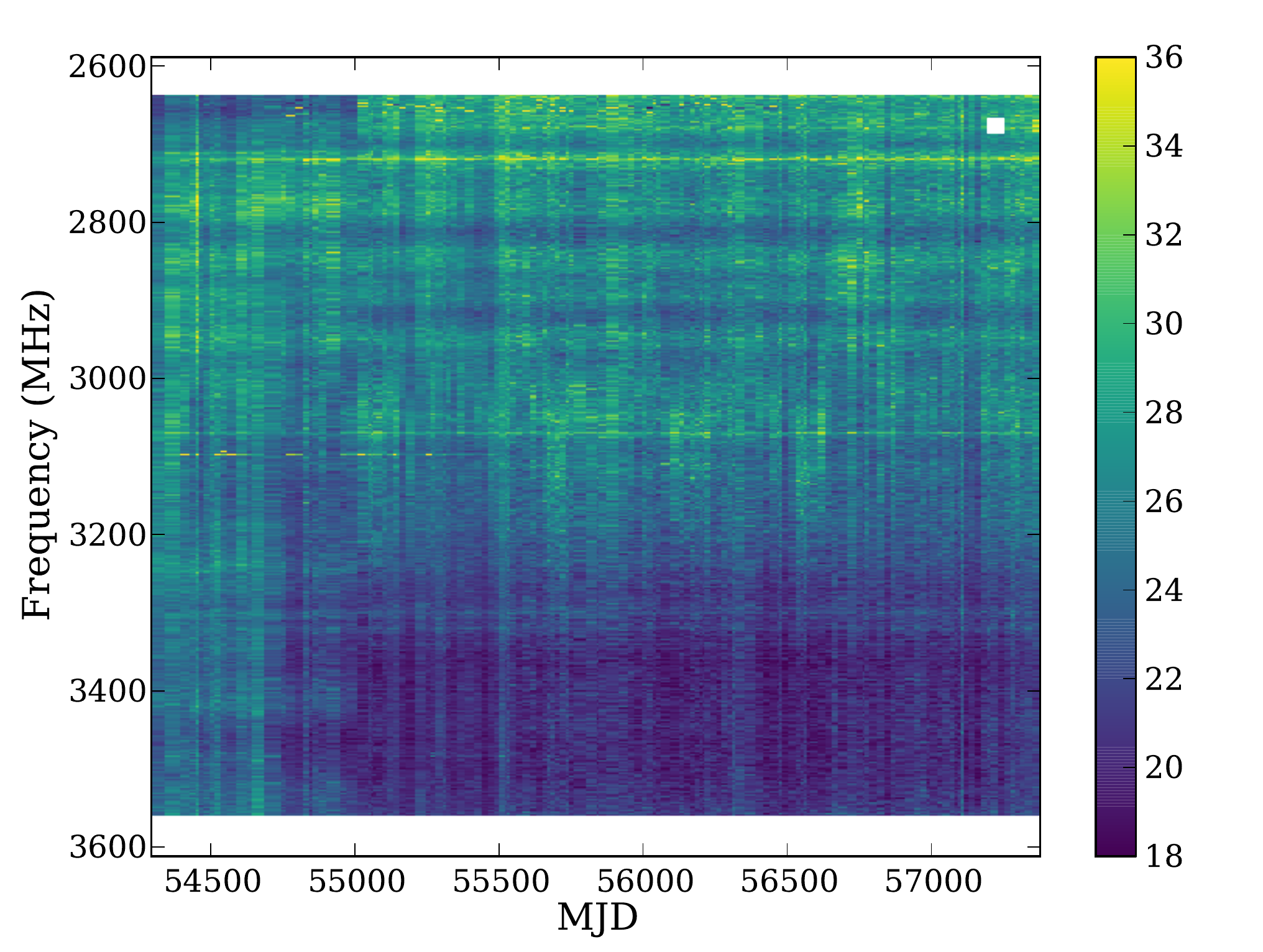}
\end{subfigure}
\begin{subfigure}[t]{0.49\textwidth}
\includegraphics[angle=0,width=\textwidth]{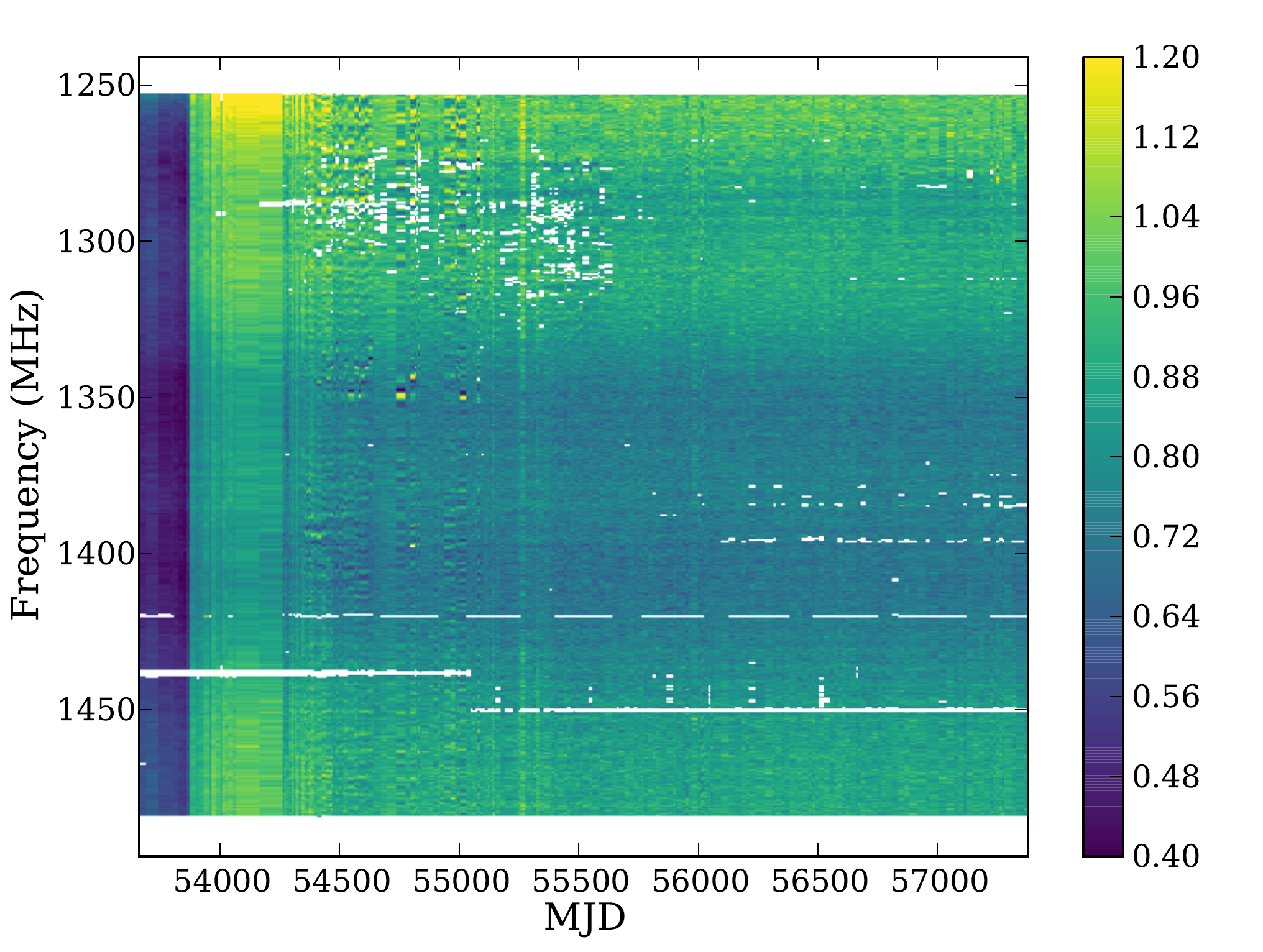}
\end{subfigure}
\begin{subfigure}[t]{0.49\textwidth}
\includegraphics[angle=0,width=\textwidth]{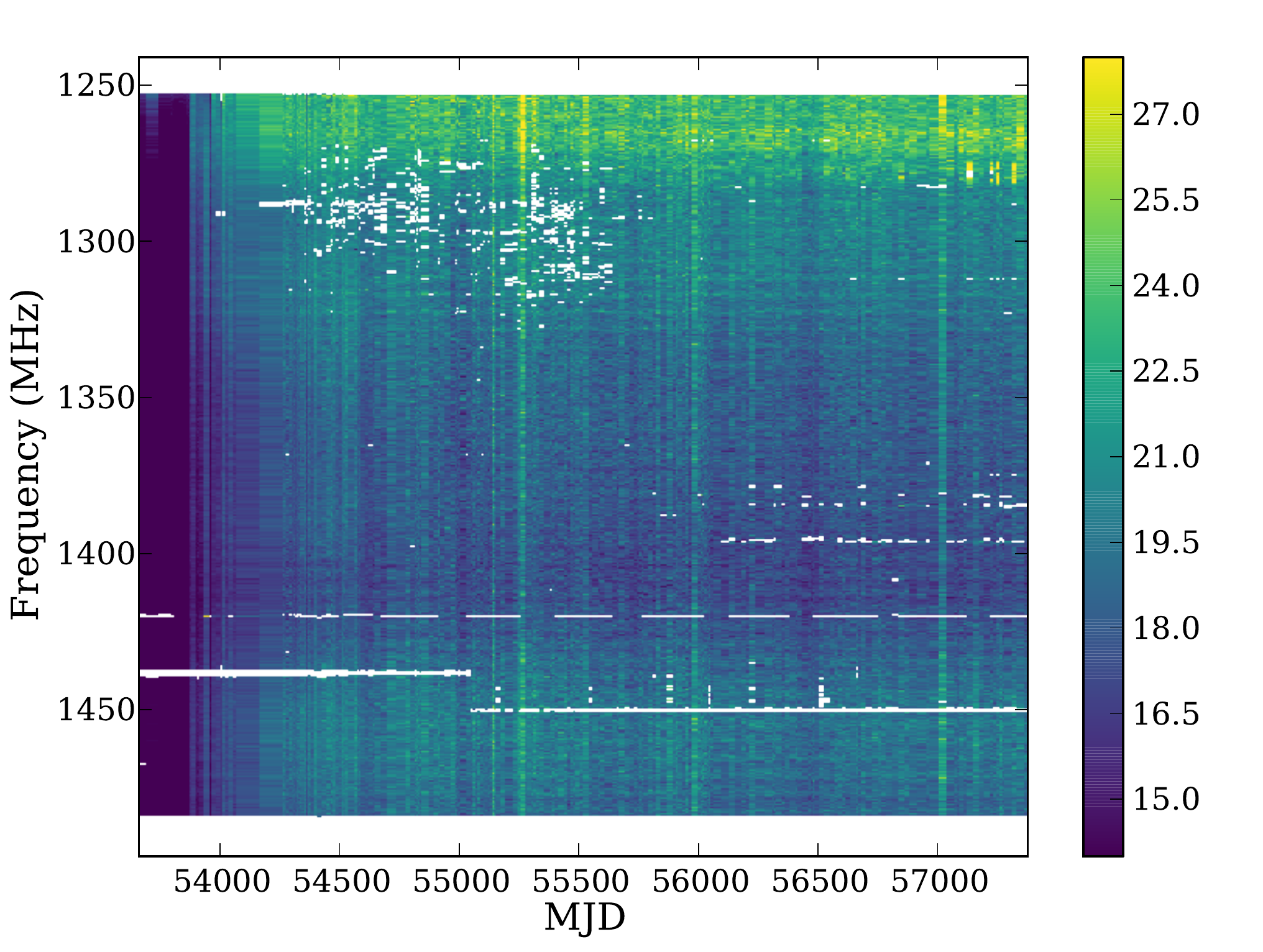}
\end{subfigure}
\begin{subfigure}[t]{0.49\textwidth}
\includegraphics[angle=0,width=\textwidth]{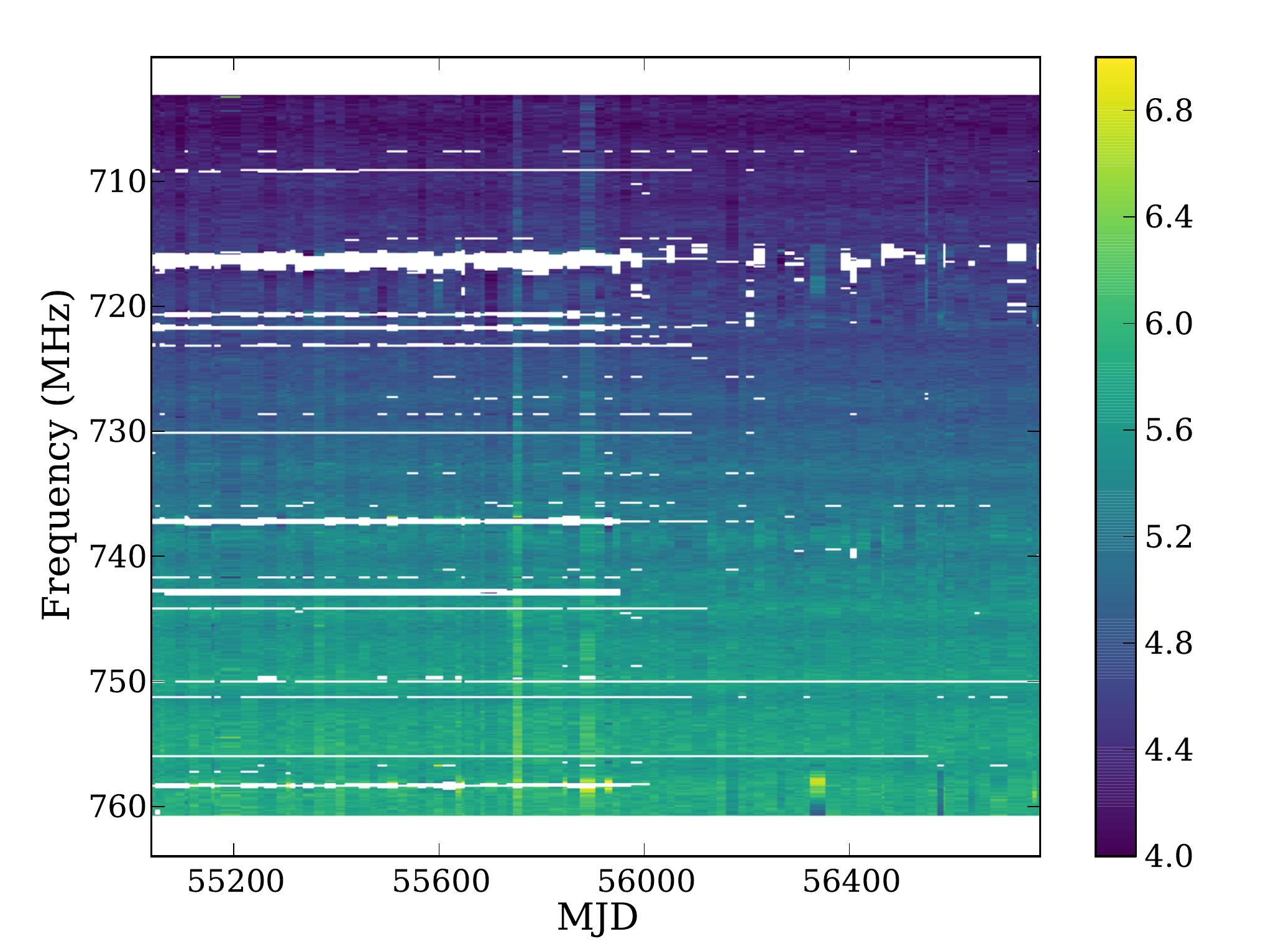}
\end{subfigure}
\begin{subfigure}[t]{0.49\textwidth}
\includegraphics[angle=0,width=\textwidth]{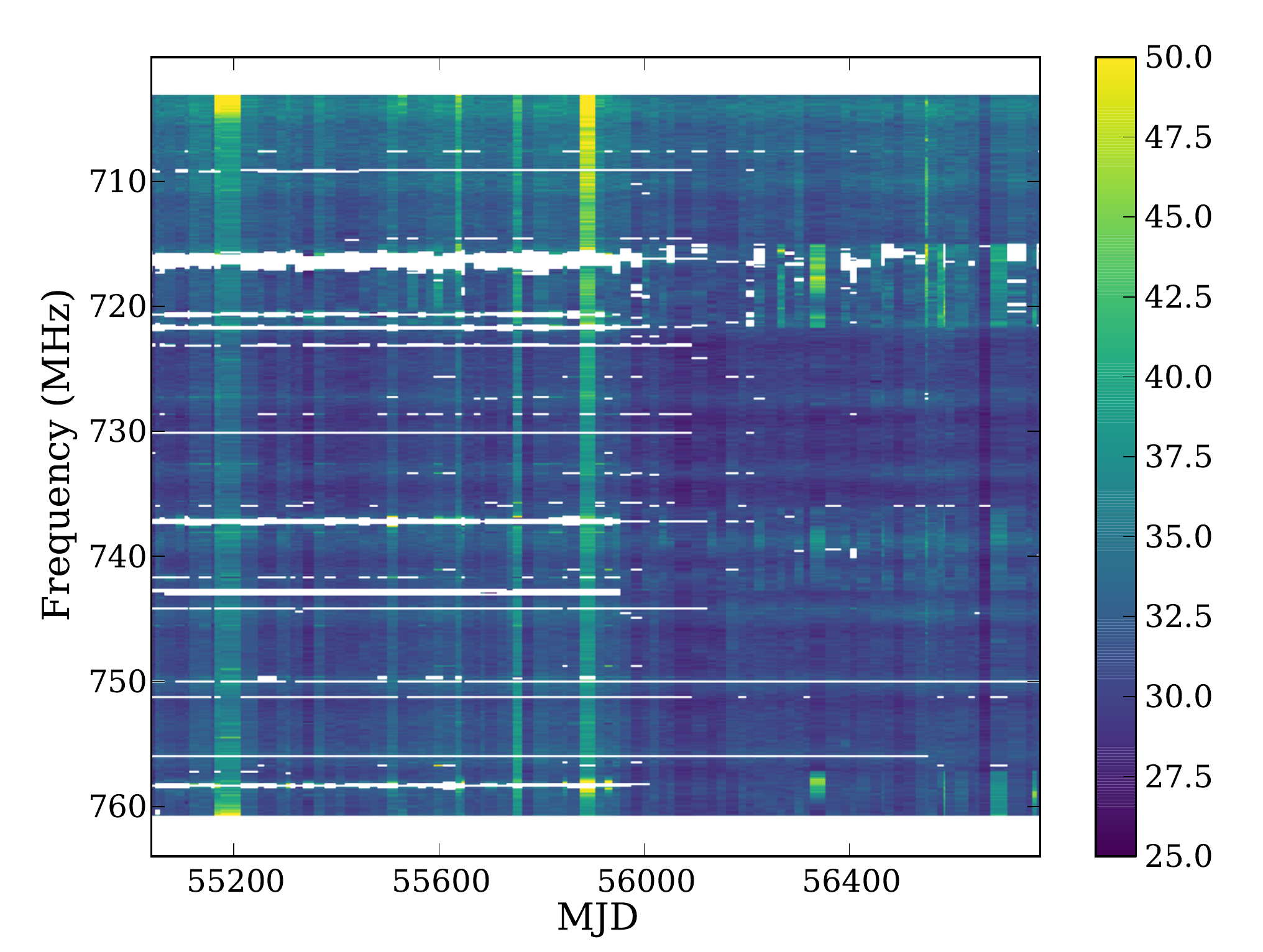}
\end{subfigure}
\caption{\label{fig:all_hydra}Variations of noise source equivalent flux density (left panels) and SEFD (right panels) as functions of time and frequency for the 10\,cm band (top), 20\,cm band (middle) and 40\,cm band (bottom) based on observations of Hydra-A. The colour bar indicates the equivalent flux density in fiducial units (Jy) in each case. Signal processors used were PDFB2 and PDFB4 for 10\,cm, PDFB1, PDFB2 and PDFB4 for 20\,cm,  and PDFB3 for 40\,cm. Individual measurements have been interpolated on to a grid for display. }
\end{figure*}

In Figure~\ref{fig:all_hydra} we provide details of how the calibration equivalent flux densities and the system equivalent flux densities (SEFD) values vary in time and frequency for the different observing systems. Step changes in SEFD spectra occurred when one or more parts of the receiver system were modified. For example, on 21 June 2006 (MJD 53907), the PDFB1 signal path was modified to improve linearity; 20\,cm SEFD measurements prior to this date are unreliable. Since the calibration source for a given system shows only modest (few per cent) long-term variations, we have averaged individual measurements over six-month intervals (unless a step change occurred during this period) to increase the signal-to-noise ratio (S/N) and to minimise small systematic errors and the number of channels given zero weight because of RFI. The specific flux density calibrator observations used for these averages are available as lists accompanying the pipeline code.

\begin{figure*}
\includegraphics[angle=-90,width=6cm]{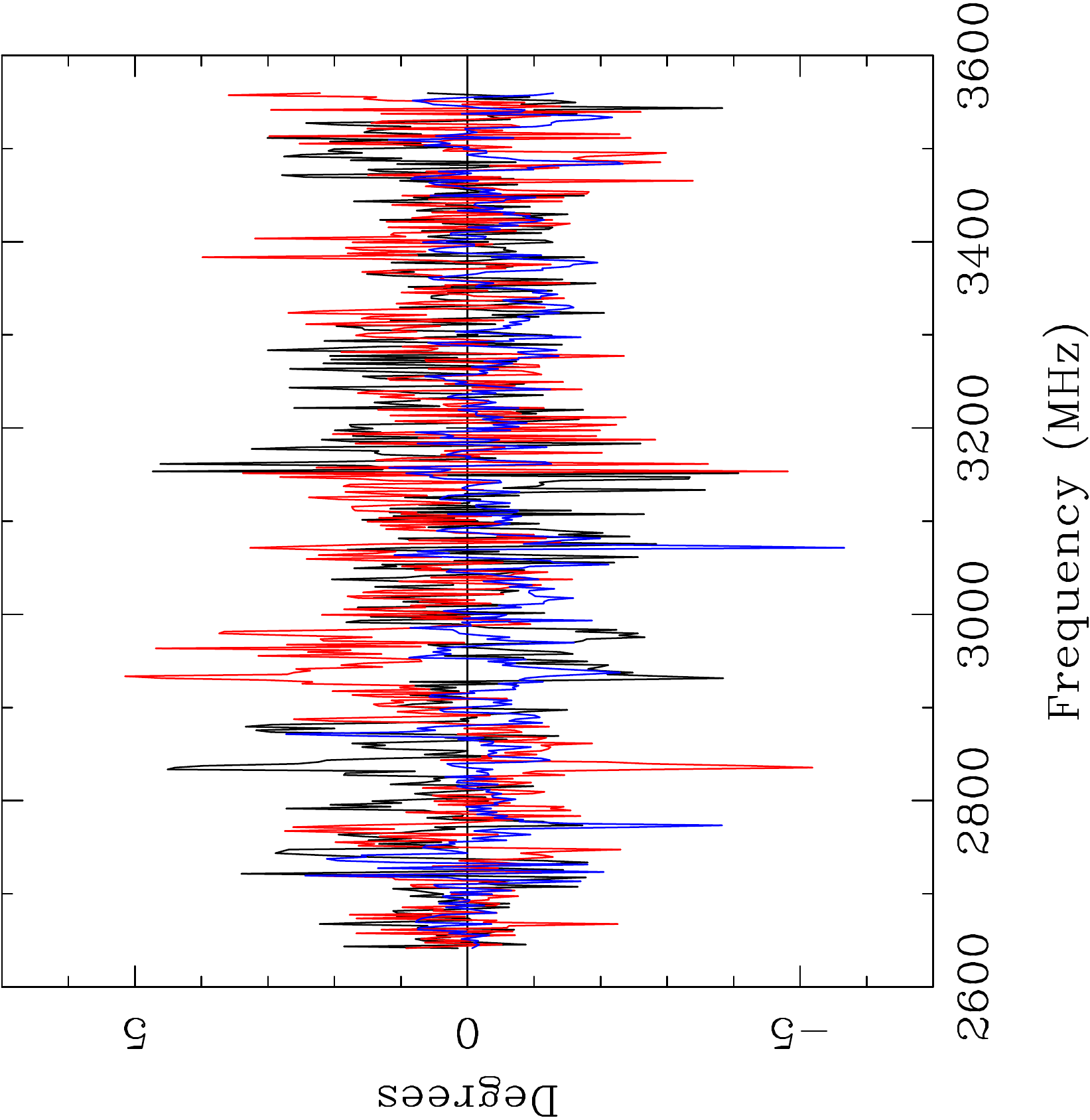}
\includegraphics[angle=-90,width=6cm]{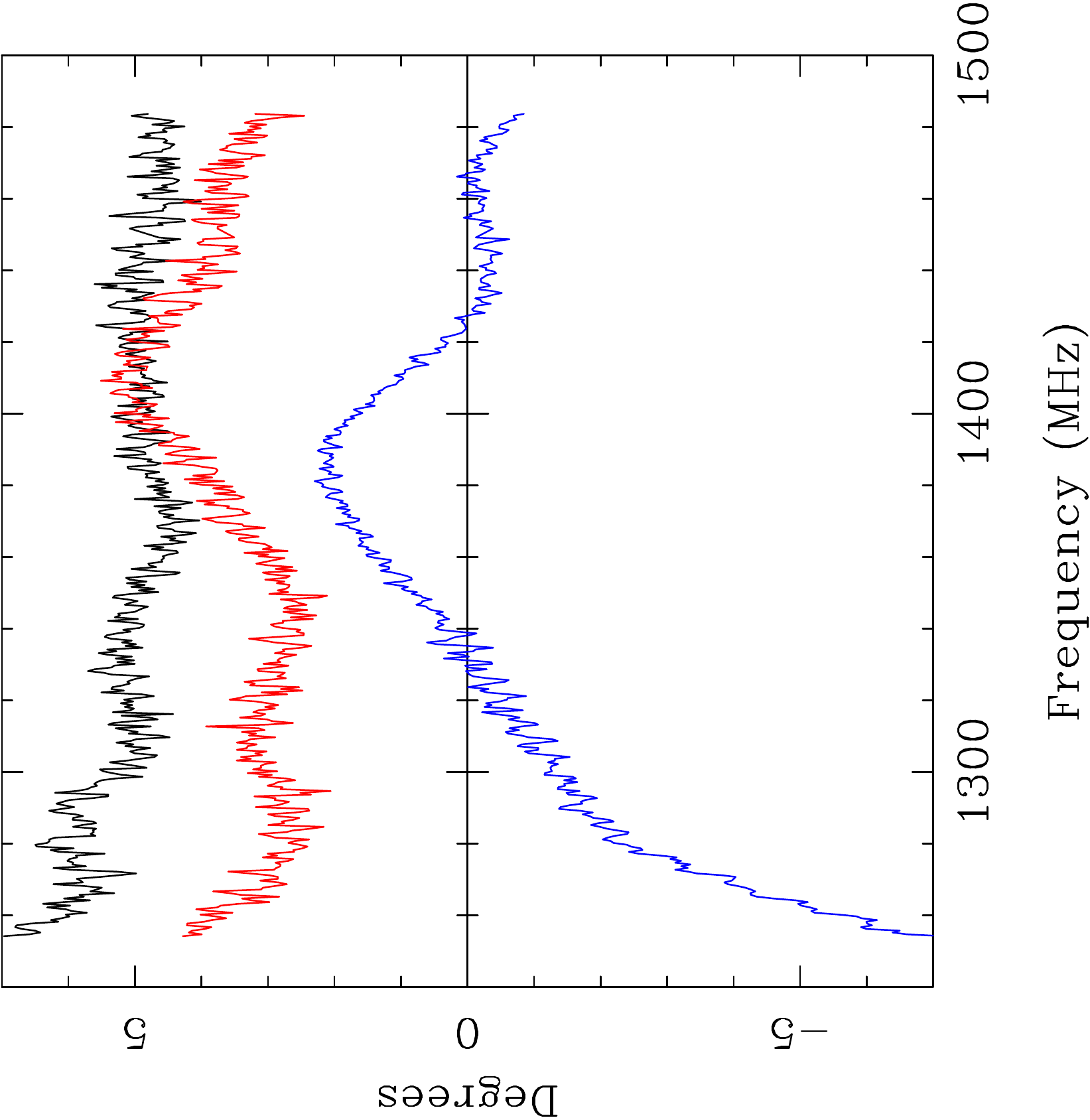}
\includegraphics[angle=-90,width=6cm]{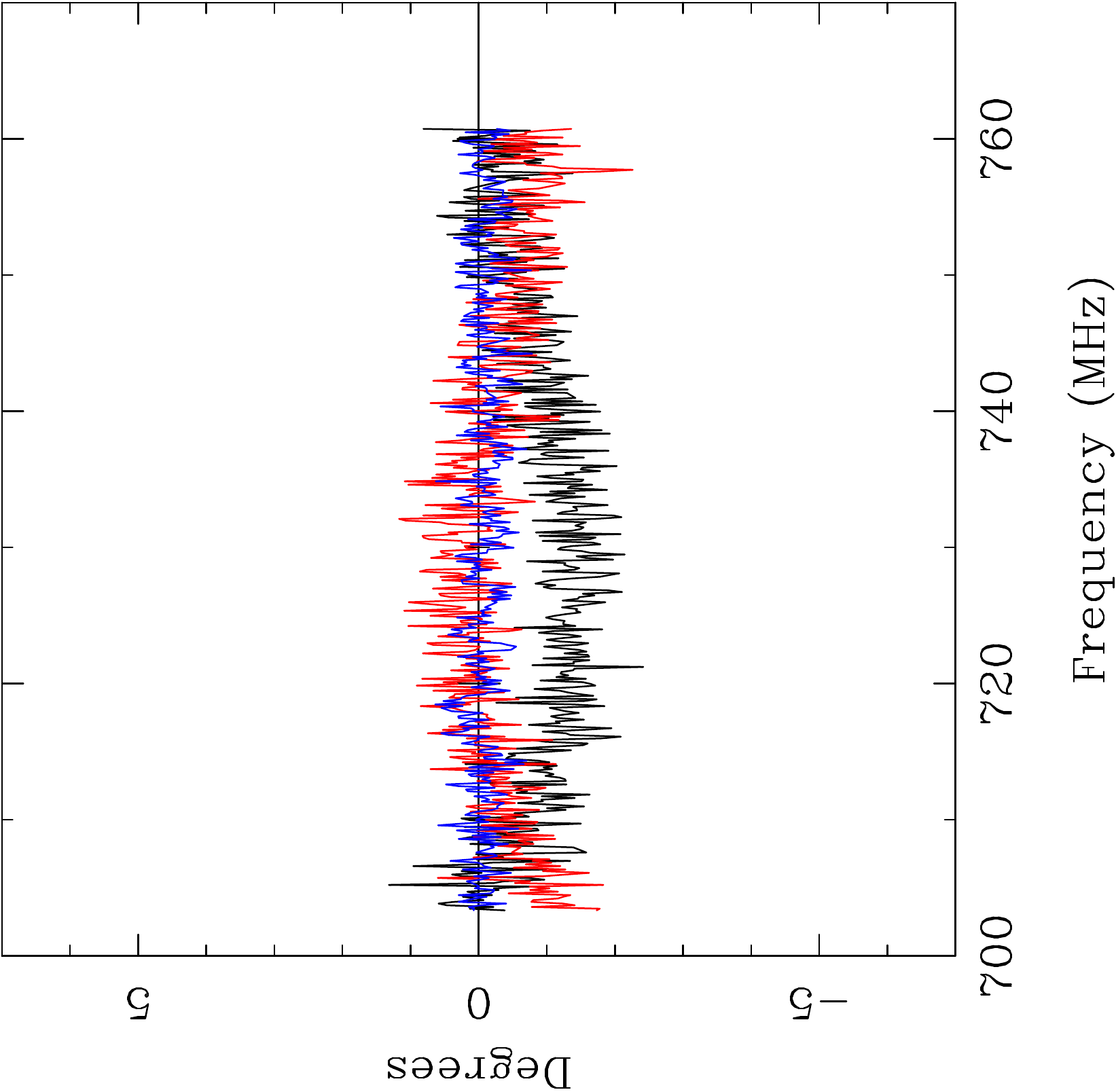}
\caption{\label{fig:pcm}Modelled receiver polarisation parameters $\epsilon_0$ (red),  $\epsilon_0$ (blue), and $\theta_1$ (black) .  Left: the 10\,cm system using PDFB4.  Centre: the 20\,cm multi-beam system using PDFB4.  Right: the 40\,cm system using PDFB3.  The H-OH receiver is not displayed because it  shows no evidence for feed cross-coupling. }
\end{figure*}

So far, we have discussed a calibration procedure sufficient for an ideal system.  Real systems depart from this ideal and may show substantial levels of cross-polarisation, feed non-orthogonality, and differential sensitivity.  We follow the method of \citet{vanStraten04} as implemented in the {\sc psrchive} routine \textit{pcm} and applied to long observations of PSR~J0437$-$4715 with most combinations of receivers and signal processors. The main results are in Figure~\ref{fig:pcm}, where we show the ellipticities of the receiver system ($\epsilon_0$ and $\epsilon_1$) as well as the orientation of receptor 1 with respect to receptor 0 ($\theta_1$). In brief, the 20 cm H-OH feed and the co-axial 40/50\,cm plus 10\,cm feed are close to the ideal over most of their bands. The 20\,cm multi-beam system shows substantial non-orthogonality and cross-polarisation.  Consequently, for the 20\,cm multibeam system we apply the full receiver model. As with the flux calibration solutions, we expect the system properties to be intrinsic to the hardware and to change only slowly, if at all, and so we combine observations to form an average solution.  For all other observing systems, we simply apply the differential gain and bandpass calibration as described above.

For PSR~J0437$-$4715, by far the brightest millisecond pulsar, and with emission extending over 90\% of the pulse period and rapidly changing polarisation properties \citep{Dai15}, we find substantial profile variation remains after polarisation calibration.  We therefore follow the same approach as in DR1 and, for PSR~J0437$-$4715, form the invariant interval \citep{bri00}, which reduces the dependence on polarisation calibration.  

Following calibration (again carried out within the \textit{psrsh} scripts), a variety of processing tasks can be performed. Typical output products include calibrated data with various levels of frequency, time, and polarisation averaging and  dynamic spectra matched-filtered with the average pulse profile. The most important for present purposes is the formation of ToA measurements for each pulsar -- this is described in detail in \S\ref{sec:dataSet} below.

\subsection{Identification and Correction of Instrumental Problems}

Over the long observing program, various systematic issues have been identified in the data set.  As a sample of such problems:
\begin{itemize}
\item Observations when digitiser levels are set incorrectly resulting in clipping or compression.
\item Observations with polarisation inputs swapped.
\item Commissioning of new instruments.
\item Periods when the observatory maser was unreliable. 
\item Observations with satellite passages through sidelobes near the primary beam.
\item Observations where the line-of-sight to the pulsar passes close to the Sun.
\end{itemize}

During the commissioning of various signal-processor systems, metadata were occasionally missing or written out incorrectly. Instances where these values are incorrect or missing have been formulated both as rule sets (e.g., date ranges with known clock failures) or as annotated individual listings of observations (e.g., for satellite passages).  During processing, these lists and rules are ingested to provide guidance for processing (e.g. zero-weight RFI-affected sub-integrations) or to propagate data quality flags to the final data products.  We note that observations with unreliable timing may still be useful for some astrophysical analysis.  

\subsection{Instrumental Timing Offsets}\label{sec:offsets}

There are generally unknown signal delays associated with each instrument.  These reflect the signal path delay and, for some instruments, the particular scheme employed for time tagging.  For example, until firmware changes on MJD~55319 were implemented, the delays of the WBCORR and PDFB instruments depended on the configuration of the filterbank (bandwidth, number of channels, number of pulse phase bins) because of the configuration-dependent delays through the FPGA processors (for the digital filterbanks). The firmware changes instead allowed time stamping at the digitiser and delays that in principle depend only on the signal path.

We have conducted a study of the delays measured and used within M+13 and found that the vast majority are accurate to the level required in our analysis (a few tens of nanoseconds).  We have also identified new delays  associated with bandpass changes and saturation effects that can affect pulse profile shapes, but they are small and difficult to separate from other sources of delay change.
   While M+13 applied corrections for these (and other) delays directly to the pulse profile metadata, we have opted to record them as \textsc{tempo2} JUMP parameters, with each observation receiving the appropriate flag(s) to trigger a JUMP/delay when appropriate.  This approach both makes the delay corrections more obvious and allows easier modification for values found to be in error.  The complete list of the fixed and floating offsets for an example pulsar is given in Appendix~\ref{sec:jumps}.

\section{The PPTA DR2 Timing Data Set}\label{sec:dataSet}

The PPTA DR2 data set contains pulse ToAs for each of the pulsars listed in Table~\ref{tb:basicParams}.
These ToAs are referenced to the local observatory time system through the time-tagging mechanisms implemented in the various signal processors. The Observatory clock and the associated 1-second pulse (1PPS) are derived from a 5\,MHz reference signal locked to a hydrogen maser frequency standard, currently a Vremya VCH-1005A unit. This 5\,MHz reference is also used to lock all local oscillators and digitiser samplers used at the Observatory. The maser is located in an air-conditioned enclosure about 100\,m east of the telescope. The phase of the 1PPS is monitored using a Global Positioning System (GPS) clock, currently a Symmetricom XL-GPS unit, located in the telescope tower, with an antenna $\sim$50\,m east of the telescope.\footnote{The propagation delay from the GPS antenna to the GPS unit is compensated for, so the GPS 1PPS is referenced to the GPS unit  location.} The offset between the Observatory 1PPS and the GPS 1PPS  is recorded at 5-minute intervals and used to form the required clock correction files. In order to maintain the Observatory 1PPS within a microsecond or so of the GPS reference, the maser is steered with rate and/or step changes. Step changes generally only occur after a system failure or transition, but rate changes to compensate for instabilities in the maser reference frequency are made more frequently, typically every few weeks. There are diurnal variations of order 10\,ns amplitude in the measured offset due to atmospheric and ionospheric variations. For pulsar timing, we use daily averages of the offset to smooth over this diurnal variation, giving an estimated accuracy in the reference time of a few nanoseconds. The clock correction chain used by {\sc tempo2} to transfer the ToA reference from the Observatory clock to TT(BIPM18) is UTC(PKS)$\rightarrow$UTC(GPS)$\rightarrow$TT(TAI)$\rightarrow$TT(BIPM18). TT(TAI)-UTC(GPS) is obtained from Circular T of the BIPM and the post-corrected time scales TT(BIPMxx)-TT(TAI), where xx is the year (20xx), are published annually by the BIPM -- see \citet{ap19} for a recent discussion of reference timescales.

We provide two arrival-time files for each pulsar: (1) a single ToA for each observation in a given receiver band where the data have been summed over frequency, and (2) multiple ToAs for sub-bands within a given band for each observation. The number of sub-bands was dynamically chosen based on the S/N of the profile and ranges from 2 to 32 for different observations and bands. Sub-banded arrival times are especially important for  pulsars  whose pulse profile evolves strongly with frequency.  Any induced variations in the centre frequency (either related to a varying RFI environment or diffractive scintillation) can add substantial measurement noise when averaging over the observing band to produce ToAs.  

Times of arrival were obtained from the calibrated profiles using a standard template for each receiver and band with the \textsc{psrchive} routine \textit{pat} implementing the \cite{tay90a} algorithm. 
We would expect to get comparable results with the monte-carlo based FDM algorithm, given the typical high S/N of the observations.
The standard templates were formed from the sum of Von Mises functions, as described in M+13, with a single frequency averaged template used for each band.  
ToAs for a given pulsar obtained with different templates will have a floating offset (a ``jump'') indicated in the corresponding parameter files. These jumps are measured with respect to PDFB4 data in the\ 10\,cm observing band.  
This band and system was chosen as the reference because it provides the highest precision timing for the best pulsars. 
The templates were approximately aligned manually and so the offsets are expected to be relatively small.   The ToAs are matched with the output from the processing pipeline to furnish estimates of the signal-to-noise ratio (S/N), goodness-of-fit, and any observation flags.  This output is incorporated directly into the \textsc{tempo2} ToA output file (see Appendix~\ref{sec:exampleToA}). 

The pipeline produces band-averaged and sub-banded ToAs for each observation. However, it is often desirable to select for or against certain groups of ToAs for particular applications. 
This is achieved through the use of the {\sc tempo2} flags associated with each ToA. For this data release we select ToAs suitable for high precision timing purposes using \textsc{tempo2} ``select'' files. An example select file is given along with further details of their use in Appendix~\ref{sec:selectFiles}.  For the data collection provided with this paper we have already applied these select files.

PPTA observations are often made with multiple signal processor systems. This is helpful in determining the instrumental time offsets between the different systems and in detection of any inconsistencies.  For the data release we provide only one set of ToAs for each observation, in general, those from the most up-to-date observing instrument.  

\begin{table*}
\caption{Observational properties of the PPTA data release. }\label{tb:obsProperties}
\begin{footnotesize}
\begin{tabular}{lrllrrrrrrrr}
\hline
PSR J & Band & Begin & End & Span & Int. Time  & $N_{\rm ToA}$ & Mean $\sigma_{\rm ToA}$ & Med $\sigma_{\rm ToA}$  & N$_{\rm ToA,sb}$ & Mean$_{\rm sb}$ & Med$_{\rm sb}$ \\
 & (cm) & (MJD) & (MJD) & (yr) &  (hr) &  & ($\mu$s) & ($\mu$s) &  & ($\mu$s) & ($\mu$s) \\
\hline
J0437$-$4715 &  10 & 53041 & 58232 & 14.2 & 1049 & 1177 & 0.04 & 0.02  & 9203 & 0.14 & 0.10  \\
 &  20 & 52741 & 58230 & 15.0 & 1427 & 1762 & 0.04 & 0.02  & 13513 & 0.16 & 0.14  \\
 &  40/50 & 52742 & 58232 & 15.0 & 1045 & 1210 & 0.16 & 0.13  & 6546 & 0.54 & 0.48  \\
\\
J0613$-$0200 &  10 & 53167 & 58209 & 13.8 & 291 & 302 & 3.22 & 2.62  & 608 & 3.71 & 3.50  \\
 &  20 & 53043 & 58230 & 14.2 & 425 & 453 & 1.01 & 0.80  & 3189 & 2.46 & 2.19  \\
 &  40/50 & 53045 & 58209 & 14.1 & 319 & 347 & 0.69 & 0.65  & 2123 & 1.82 & 1.65  \\
\\
J0711$-$6830 &  10 & 53045 & 58232 & 14.2 & 291 & 310 & 5.23 & 4.24  & 1065 & 6.54 & 6.25  \\
 &  20 & 53043 & 58230 & 14.2 & 483 & 532 & 3.15 & 2.18  & 2911 & 4.71 & 4.03  \\
 &  40/50 & 53041 & 58232 & 14.2 & 315 & 365 & 4.75 & 3.52  & 1571 & 7.37 & 6.63  \\
\\
J1017$-$7156 &  10 & 55459 & 58229 & 7.6 & 169 & 161 & 1.88 & 1.82  & 329 & 2.58 & 2.02  \\
 &  20 & 55394 & 58231 & 7.8 & 332 & 340 & 0.44 & 0.41  & 2596 & 1.11 & 1.02  \\
 &  40/50 & 55427 & 58232 & 7.7 & 233 & 233 & 0.87 & 0.81  & 1128 & 2.05 & 2.17  \\
\\
J1022+1001 &  10 & 53045 & 58209 & 14.1 & 289 & 308 & 1.80 & 1.40  & 2386 & 4.81 & 4.01  \\
 &  20 & 53048 & 58230 & 14.2 & 445 & 525 & 2.08 & 1.23  & 3748 & 4.22 & 3.08  \\
 &  40/50 & 53042 & 58209 & 14.1 & 274 & 297 & 3.11 & 1.88  & 1522 & 4.98 & 3.85  \\
\\
J1024$-$0719 &  10 & 53083 & 58021 & 13.5 & 201 & 210 & 4.98 & 4.64  & 361 & 5.67 & 5.34  \\
 &  20 & 53043 & 58187 & 14.1 & 318 & 337 & 2.54 & 1.75  & 1740 & 3.50 & 3.23  \\
 &  40/50 & 53042 & 57857 & 13.2 & 151 & 159 & 4.26 & 3.67  & 542 & 4.81 & 4.67  \\
\\
J1045$-$4509 &  10 & 53518 & 58021 & 12.3 & 242 & 245 & 9.01 & 8.10  & 770 & 14.04 & 13.45  \\
 &  20 & 53043 & 58212 & 14.2 & 377 & 397 & 2.84 & 2.05  & 3003 & 7.29 & 5.84  \\
 &  40/50 & 53046 & 58021 & 13.6 & 260 & 284 & 3.49 & 2.83  & 1838 & 8.23 & 7.00  \\
\\
J1125$-$6014 &  10 & 56892 & 58230 & 3.7 & 75 & 78 & 0.87 & 0.71  & 247 & 1.03 & 0.99  \\
 &  20 & 53723 & 58229 & 12.3 & 130 & 191 & 0.97 & 0.75  & 989 & 1.67 & 1.52  \\
 &  40/50 & 56868 & 58230 & 3.7 & 76 & 81 & 1.35 & 1.17  & 171 & 1.60 & 1.51  \\
\\
J1446$-$4701 &  10 & 56137 & 58210 & 5.7 & 17 & 16 & 1.45 & 1.40  & 39 & 1.60 & 1.61  \\
 &  20 & 55521 & 58211 & 7.4 & 137 & 149 & 1.55 & 1.36  & 368 & 1.93 & 1.81  \\
 &  40/50 & 55594 & 58210 & 7.2 & 71 & 72 & 2.66 & 2.09  & 101 & 3.09 & 2.31  \\
\\
J1545$-$4550 &  10 & 55691 & 58233 & 7.0 & 100 & 103 & 1.04 & 0.94  & 441 & 1.80 & 1.74  \\
 &  20 & 55685 & 58231 & 7.0 & 155 & 167 & 1.03 & 0.87  & 1123 & 2.36 & 2.24  \\
 &  40/50 & 55988 & 58229 & 6.1 & 80 & 78 & 5.03 & 4.26  & 70 & 4.75 & 4.09  \\
\\
J1600$-$3053 &  10 & 53041 & 58233 & 14.2 & 265 & 287 & 0.90 & 0.69  & 2053 & 2.27 & 1.97  \\
 &  20 & 53043 & 58232 & 14.2 & 441 & 540 & 0.59 & 0.48  & 3953 & 1.57 & 1.26  \\
 &  40/50 & 53041 & 58233 & 14.2 & 257 & 269 & 1.93 & 1.65  & 1041 & 3.53 & 3.46  \\
\\
J1603$-$7202 &  10 & 53046 & 58233 & 14.2 & 200 & 204 & 6.47 & 5.80  & 518 & 7.31 & 7.20  \\
 &  20 & 53087 & 58232 & 14.1 & 386 & 412 & 1.57 & 1.01  & 3071 & 3.84 & 2.90  \\
 &  40/50 & 53042 & 58233 & 14.2 & 244 & 275 & 2.23 & 1.70  & 1758 & 5.41 & 4.37  \\
\\
J1643$-$1224 &  10 & 53045 & 58233 & 14.2 & 235 & 248 & 1.85 & 1.47  & 1805 & 4.70 & 4.17  \\
 &  20 & 53043 & 58232 & 14.2 & 299 & 326 & 0.97 & 0.68  & 2559 & 2.66 & 1.94  \\
 &  40/50 & 53084 & 58233 & 14.1 & 228 & 244 & 1.74 & 1.38  & 1577 & 4.10 & 3.38  \\
\\
J1713+0747 &  10 & 53041 & 58233 & 14.2 & 282 & 291 & 0.32 & 0.23  & 2298 & 0.96 & 0.66  \\
 &  20 & 53043 & 58232 & 14.2 & 429 & 479 & 0.23 & 0.12  & 3741 & 0.81 & 0.39  \\
 &  40/50 & 53041 & 58233 & 14.2 & 271 & 279 & 1.11 & 0.86  & 1765 & 3.63 & 2.46  \\
\\
J1730$-$2304 &  10 & 53045 & 58233 & 14.2 & 198 & 213 & 4.42 & 2.80  & 1020 & 5.87 & 5.24  \\
 &  20 & 53043 & 58230 & 14.2 & 267 & 304 & 1.85 & 1.11  & 2156 & 4.29 & 3.34  \\
 &  40/50 & 53041 & 58233 & 14.2 & 195 & 224 & 2.41 & 1.85  & 1373 & 5.44 & 4.68  \\
\\
J1732$-$5049 &  10 & 53084 & 55725 & 7.2 & 49 & 50 & 8.39 & 7.48  & 85 & 9.55 & 8.76  \\
 &  20 & 53725 & 55724 & 5.5 & 77 & 85 & 2.08 & 1.95  & 658 & 6.44 & 5.65  \\
 &  40/50 & 55041 & 55582 & 1.5 & 8 & 8 & 3.06 & 2.77  & 64 & 7.96 & 7.81  \\
\hline
\end{tabular}
\end{footnotesize}
\end{table*}
\begin{table*}
\caption{(Continued) Observational properties of the PPTA data release.}
\begin{footnotesize}
\begin{tabular}{lrllrrrrrrrr}
\hline
PSR J & Band & Begin & End & Span &  Int. Time & $N_{\rm ToA,sb}$ & Mean $\sigma_{\rm ToA}$ & Med $\sigma_{\rm ToA}$ & N$_{\rm ToA,sb}$ & Mean$_{\rm sb}$ & Med$_{\rm sb}$ \\
 & (cm) & (MJD) & (MJD) & (yr) &  (hr) & & ($\mu$s)  & ($\mu$s) &  & ($\mu$s) & ($\mu$s) \\
\hline
J1744$-$1134 &  10 & 53041 & 58233 & 14.2 & 263 & 277 & 0.99 & 0.76  & 1736 & 1.92 & 1.74  \\
 &  20 & 53043 & 58232 & 14.2 & 402 & 457 & 0.65 & 0.38  & 3289 & 1.74 & 1.02  \\
 &  40/50 & 53041 & 58233 & 14.2 & 274 & 285 & 0.82 & 0.58  & 1692 & 2.37 & 1.50  \\
\\
J1824$-$2452A &  10 & 53521 & 58203 & 12.8 & 120 & 127 & 1.52 & 1.26  & 254 & 1.89 & 1.76  \\
 &  20 & 53519 & 58203 & 12.8 & 204 & 218 & 0.59 & 0.43  & 1620 & 1.46 & 1.14  \\
 &  40/50 & 53164 & 58177 & 13.7 & 122 & 130 & 1.43 & 1.18  & 752 & 2.88 & 2.65  \\
\\
J1832$-$0836 &  10 & 56497 & 57783 & 3.5 & 23 & 24 & 0.91 & 0.73  & 60 & 1.24 & 1.23  \\
 &  20 & 56260 & 58232 & 5.4 & 59 & 65 & 0.84 & 0.75  & 241 & 1.38 & 1.32  \\
 &  40/50 & 56497 & 57783 & 3.5 & 21 & 22 & 2.36 & 2.24  & 25 & 2.46 & 2.34  \\
\\
J1857+0943 &  10 & 53042 & 58210 & 14.1 & 94 & 183 & 3.87 & 3.04  & 644 & 4.69 & 4.35  \\
 &  20 & 53087 & 58230 & 14.1 & 148 & 281 & 1.38 & 1.13  & 2069 & 4.08 & 3.31  \\
 &  40/50 & 53042 & 58233 & 14.2 & 100 & 199 & 3.11 & 2.76  & 1127 & 7.07 & 6.54  \\
\\
J1909$-$3744 &  10 & 53041 & 58221 & 14.2 & 615 & 635 & 0.21 & 0.14  & 4573 & 0.45 & 0.35  \\
 &  20 & 53043 & 58230 & 14.2 & 790 & 976 & 0.27 & 0.15  & 6632 & 0.61 & 0.38  \\
 &  40/50 & 53041 & 58221 & 14.2 & 589 & 622 & 0.31 & 0.23  & 3422 & 0.90 & 0.63  \\
\\
J1939+2134 &  10 & 53084 & 58210 & 14.0 & 95 & 182 & 0.38 & 0.20  & 1268 & 0.70 & 0.55  \\
 &  20 & 53083 & 58230 & 14.1 & 152 & 298 & 0.07 & 0.06  & 2360 & 0.24 & 0.20  \\
 &  40/50 & 53084 & 58210 & 14.0 & 96 & 190 & 0.12 & 0.08  & 1313 & 0.42 & 0.37  \\
\\
J2124$-$3358 &  10 & 53045 & 58209 & 14.1 & 131 & 255 & 8.92 & 8.04  & 360 & 10.03 & 9.69  \\
 &  20 & 53043 & 58230 & 14.2 & 263 & 518 & 3.20 & 2.38  & 3517 & 6.47 & 5.95  \\
 &  40/50 & 53041 & 58209 & 14.1 & 123 & 245 & 4.69 & 3.91  & 1064 & 6.09 & 5.44  \\
\\
J2129$-$5721 &  10 & 53881 & 58095 & 11.5 & 12 & 24 & 6.07 & 6.02  & 28 & 5.91 & 6.00  \\
 &  20 & 53205 & 58232 & 13.8 & 354 & 363 & 2.36 & 1.86  & 1635 & 3.38 & 3.15  \\
 &  40/50 & 53163 & 58130 & 13.6 & 131 & 254 & 1.70 & 1.39  & 1216 & 3.29 & 2.91  \\
\\
J2145$-$0750 &  10 & 53084 & 58209 & 14.0 & 267 & 280 & 2.17 & 1.62  & 2105 & 5.23 & 4.54  \\
 &  20 & 53110 & 58230 & 14.0 & 392 & 447 & 2.08 & 0.97  & 3125 & 3.71 & 2.33  \\
 &  40/50 & 53084 & 58130 & 13.8 & 252 & 267 & 2.23 & 1.19  & 1637 & 4.77 & 3.13  \\
\\
J2241$-$5236 &  10 & 55236 & 58220 & 8.2 & 212 & 218 & 0.67 & 0.60  & 1061 & 1.33 & 1.28  \\
 &  20 & 55235 & 58230 & 8.2 & 374 & 408 & 0.21 & 0.16  & 3127 & 0.55 & 0.42  \\
 &  40/50 & 55236 & 58220 & 8.2 & 188 & 195 & 0.50 & 0.25  & 1036 & 0.84 & 0.48  \\
\\
\hline
\end{tabular}
\end{footnotesize}
Note: Int. Time is the total integration time.   $N_{ToA,sb}$ is the number of sub-banded ToAs in each data set. 
\end{table*}

\begin{figure*}
    \centering
    \includegraphics[width=15cm]{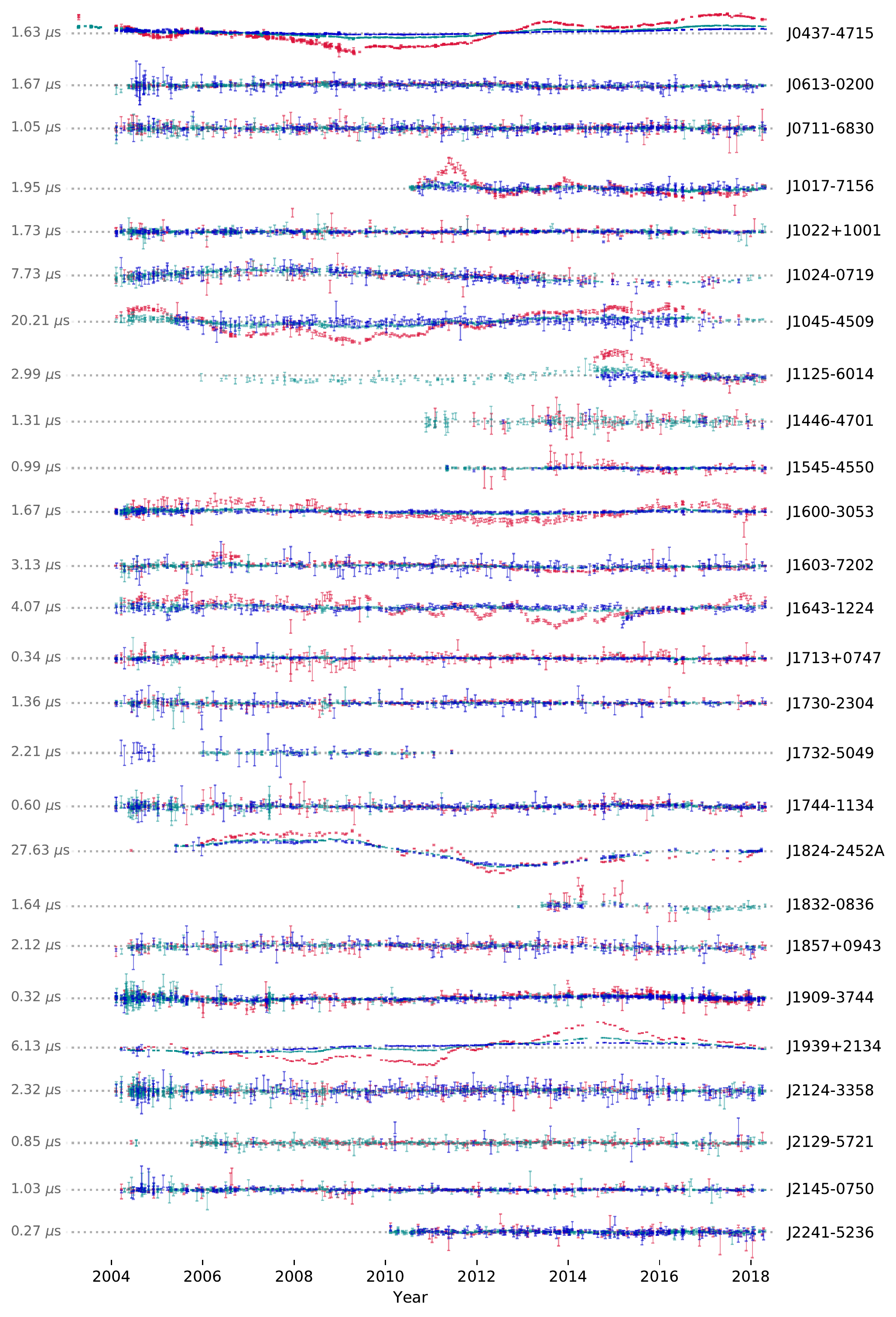}
    \caption{Band-averaged timing residuals for the PPTA DR2 pulsars prior to fitting for DM variations and frequency-independent red noise. The value to the left of the residuals for each pulsar is the weighted rms residual from the model fit. The 10\,cm, 20\,cm and 40\,cm observing bands are shown in blue, cyan, and red respectively. }
    \label{fg:timingResiduals}
\end{figure*}

\begin{figure*}
    \centering
    \includegraphics[width=15cm]{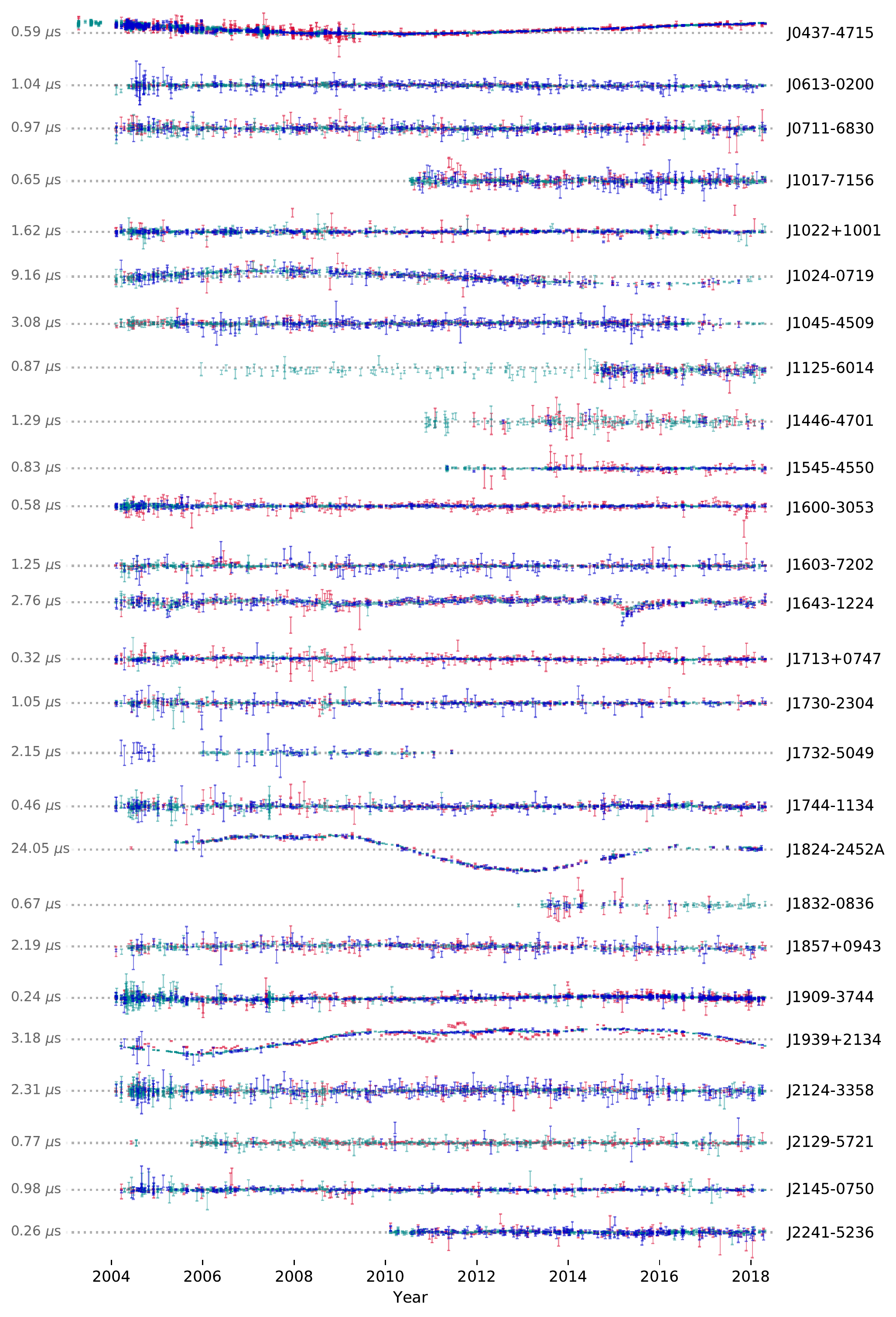}
    \caption{Timing residuals for the PPTA DR2 pulsars as in Figure~\ref{fg:timingResiduals}, but with the best-fitting realisation of the DM variations subtracted.}
    \label{fg:timingResiduals_subdm}
\end{figure*}

Table~\ref{tb:obsProperties} lists the pulsar name, observing band, MJDs of the first and last observations, the overall data span, the total integration time, the number of ToAs in the band-averaged data and the mean and median ToA uncertainties (as directly measured from the observations without noise modelling). The final three columns provide the number of sub-banded ToAs and their corresponding mean and median uncertainty.   

For each pulsar we determine initial noise models using the \textsc{Enterprise}\footnote{\url{https://github.com/nanograv/enterprise}} package \citep{enterprise}, to account for any low-frequency (``red'') noise in the residuals. These noise models include a simple power law for frequency-independent red noise, such as intrinsic spin variations, as well as a frequency-dependent power law with $\nu^{-2}$ scaling to account for dispersion measure variations, where $\nu$ is the radio frequency. These power laws were fitted to the data as the amplitudes of sine-cosine Fourier component pairs, with frequencies $k/T_{\rm span}$, where $k=1,\dots,N$ for $N$ components, and $T_{\rm span}$ is the total observing span. The number of Fourier components $N$ was chosen for each pulsar such that the highest frequency modelled, $N/T_{\rm span}$ was equivalent to $1/60$\, days. Lower frequency red noise is absorbed into the spin frequency and spin-down parameters. Note that the first and second time derivatives of DM were also included in the timing model for all pulsars, in order to pre-whiten any DM variations.  Additional band-dependent noise models  were required for PSRs J0437$-$4715 and J1939+2134  \citep[see][]{Lentati16}. The exponential DM events observed in PSR~J1713$+$0747 \citep{lam+18}, and the magnetospheric event observed in PSR~J1643$-$1224 \citep[shown in Figure \ref{fg:timingResiduals_subdm};][]{slk+16} were modelled as chromatic exponential functions using \textsc{Enterprise} simultaneously with noise parameters, and then included into the timing model using extra parameters.  The contribution to the timing residuals is modelled to be 
\begin{equation}
R(t) = \theta(t-t_0) A \left( \frac{\nu}{1.4\,{\rm GHz}} \right)^\alpha \exp \left( - \frac{t-t_0}
{\tau} \right),
\end{equation}
where $\theta(t)$ is the Heaviside (step) function, $t_0$ is the epoch of the event, $A$ its amplitude in $\mu$s, $\tau$ is the decay time scale in days, and $\alpha$ is the spectral index of the frequency scaling (e.g. $\alpha =-2$ for DM-like events and $\alpha =0$ for achromatic events).  
The exponential fitting function is chosen as it phenomenologically matches the function in the arrival times \citep{slk+16,EPTA}. 
These fitting functions have been incorporated into the {\sc tempo2} repository.

To account for pulse profile evolution with radio frequency, our parameter files also include ``FD'' parameters for all pulsars except PSRs~J1446$-$4701 and J1832$-$0836. These parameters are the coefficients of a polynomial function fitted to the logarithm of observing frequency \citep{Arzoumanian15}.   The number of FD parameters was determined using the Akaike information criterion \citep[AIC;][]{Akaike73}. 

For PSR~J0437$-$4715, we identified an inconsistency in the apparent pulse profile evolution for some early observing systems (CPSR2, PDFB1, and WBCORR) relative to newer systems.  We attribute this to  frequency- and phase-dependent sensitivity changes, rather than intrinsic profile evolution. These system-induced profile evolution components are most significant for PSR~J0437$-$4715 because of its brightness. 
For this pulsar we measured the FD parameters relative to stable system (PDFB3, PDFB4, and CASPSR), for all observing bands. The excess apparent profile evolution in the early systems was then measured relative to these by fitting a polynomial (with order up to a cubic function, as determined by the AIC), as a function of frequency.  This offset was then subtracted directly from the ToAs.  

For PSR~J2241$-$5236 we also included six orbital frequency derivatives in the timing model to account for observed variations in the orbital period.
These additional parameters are not expected to significantly reduce the sensitivity of the observations to low-frequency gravitational waves \cite[][]{Bochenek15}. 
We note that a detailed analysis of noise models for DR2 pulsars will be presented elsewhere.  These initial models are used here simply to ensure the integrity of our ToAs, and to allow for initial fitting of the system jump parameters. 

For the figures described below we have formed residuals from the sub-banded ToAs, and then averaged (with weighting) the resulting residuals for each observing system for each observation.
This enables us to account for profile evolution in the fitting, but also to see weak signals that would be difficult to identify in a single sub-banded observation. 
In Figure~\ref{fg:timingResiduals} we show band-averaged timing residuals for the PPTA DR2 pulsars before fitting for DM variations and frequency-independent red noise.  For many pulsars, there are large systematic residuals resulting from DM variations and/or frequency independent timing noise. The quoted weighted rms residuals reflect these systematic variations. Note that an apparent offset in the low frequency (red, 40\,cm and 50\,cm) residuals for PSR~J0437$-$4715 corresponds to the mid-2009 change in the centre frequency of this band, and is caused by the $\nu^{-2}$ scaling of the DM variations.

 Timing residuals as in Figure~\ref{fg:timingResiduals},   but after the best-fitting DM variations have been removed, are shown in Figure~\ref{fg:timingResiduals_subdm}.  This figure highlights the frequency-independent timing noise present in the residuals for some pulsars, e.g., PSR~J1824$-$2452A. However, band-dependent noise still appears to be present in some pulsars, e.g., PSR~J1939+2134, where the 40\,cm residuals show some features not accounted for by DM modelling. These deviations may result from changes in the amount of interstellar scattering modifying the 40\,cm profile shape \citep{coles15}.   
 We note the fitting incorporates a generalized least square algorithm to  account for both dispersion measure variations and achromatic timing noise. This fitting method will reveal quadratic features in the timing residuals that would be absorbed in a simple weighted least squares fit.

Our data collection contains the pulse arrival times for each pulsar along with our initial timing ephemerides and noise models. We also provide data files used during the pipeline processing including pulse profiles, standard templates, the flux and polarisation calibration files and the \textit{psrsh} scripts.  Full details are provided in the README file provided with the data collection.

\section{Discussion}\label{sec:discussion}

\subsection{The DR2 Data Set}

The PPTA DR2 data set significantly improves on our DR1 data set \citep{Manchester13}. Data spans are significantly longer with an extra eight years of data from 2011 to 2018. For PSR~J1909$-$3744, one of the best-timed pulsars, the data span in DR2 is now twice that of DR1. For most pulsars, the data span is about 14\,yr, which is greater than the orbital period of Jupiter. This will improve our ability to detect errors in the solar system ephemerides, which are already known to affect pulsar timing array data sets  \citep{Tiburzi16,Nanograv11GWB}. 

In contrast to the first data release, DR2 provides ToAs for sub-bands within the bandwidth of each instrument as well as band-averaged ToAs. There are advantages and disadvantages to each type of data set. For instance, the sub-band files for PSR~J0437$-$4715 contain 29,316 ToAs. Processing such large numbers of ToAs is computationally expensive (in particular, for Bayesian inference and Monte-Carlo simulations) and small variations in the residuals are difficult to identify by eye.    In contrast, the sub-banded ToAs enable studies of pulse shape evolution, detailed studies of interstellar medium effects, greater finesse in flagging observations, and a more accurate assessment of the scatter in the residuals.  Sub-banded ToAs are especially valuable where the mean pulse profile changes over the band and broad-band scintillation is present.  Consequently, the choice of whether to use the sub-band ToAs or the band-averaged ToAs depends on the goal of the data processing, on the particular pulsar and on the computing power available.

The DR2 arrival times can be used for many purposes, but we caution that the modelling of timing noise and instrumental jumps is preliminary and should be reassessed for any particular application. To aid in this we have provided initial pulsar emphemerides containing pulsar parameters, instrumental jumps and noise models. As some of the jumps, pulsar parameters and noise models are covariant, we have provided initial Bayesian noise parameters.  These can be used with an up-to-date version of \textsc{tempo2}.\footnote{\url{https://bitbucket.org/psrsoft/tempo2}}  

\subsection{Known timing events affecting the DR2 pulsars}

The following timing events have already been detected in the timing residuals for the DR2 pulsar sample:

\begin{itemize}
    \item PSR~J0437$-$4715: In 2015 Feb., a variation in the profile for this pulsar was detected. We will analyse and present details of this event elsewhere.
    \item PSR~J1017$-$7156: An extreme scattering event (ESE) was reported between MJDs 55600 and 55800 by \citet{coles15}.
    \item PSR~J1024$-$0719 is thought to be in a long-period binary orbit; see \citet{bjs+16} and \citet{kkn+16}.
    \item PSR~J1603$-$7202: An ESE between MJDs 53740 and 54000 was identified by \citet{coles15}.
    \item PSR~J1643$-$1224: \citet{slk+16} reported on profile   variations starting in 2015 March. This event can easily be seen by eye in Figure~\ref{fg:timingResiduals}. 
    \item PSR~J1713$+$0747: \citet{coles15}, \cite{EPTA}, and \citet{Lentati16} showed that the line-of-sight to this pulsar went through an under-dense region of the interstellar medium around 2008 November (MJD~54800). A second similar event at MJD~57500 was reported by \citep{lam+18}. These events are too small to be easily seen in Figure~\ref{fg:timingResiduals}
    \item PSRs J1022$+$1002, J1730$-$2304, and J1824$-$2452A are located near the ecliptic plane and show significant excess dispersive delay from the solar wind when close to the Sun on the sky.   While observations at these epochs are useful for studying the structure solar wind, care would need taken before including them in precision-timing studies \cite[][]{Tiburzi19}.
\end{itemize}

\subsection{Continuing the PPTA observations}

The primary goal for PTA projects is to detect and study ultra-low-frequency gravitational waves.  Bounds of the amplitude of any such  gravitational waves are currently dominated by a few of the best-timed pulsars (e.g., PSRs~J0437$-$4715, J1909$-$3744 and J1713$+$0747) \citep[see, e.g.,][]{Shannon15}.  However, any detection of a GW background signal would require strong evidence for a measurement of the ``Hellings-Downs'' curve that describes the signal correlation between pulsars as a function of their angular separation \citep{hd83,Tiburzi16}. This requires a large sample of pulsars that are timed over long data spans \citep{siemens+13}. 

However, it is doubtful that pulsars such as PSRs~J1939$+$2134 or J1824$-$2452A, which exhibit significant timing noise, will contribute to the detection of a gravitational-wave background. We continue to observe such pulsars for the following reasons:
\begin{itemize}
\item The observations are intrinsically interesting, allowing studies of e.g. binary evolution, globular cluster dynamics, and variations in the interstellar medium. All of these also represent noise sources for precision timing, so their study can help improve timing precision.
\item PSR~J1939+2134, as the first discovered millisecond pulsar \citep{bkh+82} provides the longest data spans available for millisecond pulsar timing.
It is a very strong pulsar and provides some of the most precise ToAs measured. Although its red noise limits its value for detecting long period GWs, its short term precision makes it valuable for detecting events such as gravitational waves from: memory-events; cosmic strings; bursts; sinusoidal sources with periods less than a few years; or periastron passages. Since it is also visible to all IPTA telescopes it is valuable for synchronizing timing, although care must be taken when the frequencies are different as it shows excess chromatic noise due to scattering in the ISM \citep{Lentati16}. 
\item The detection of gravitational-wave burst events (either memory events, burst signatures or periastron passages) does not necessarily require long data spans that are unaffected by timing noise.  For instance, PSR~J1939$+$2134 provides some of the most precise ToAs measured.  If a burst event occurs over relatively short timescales then such pulsars will be invaluable in detecting and studying the burst.  For similar reasons, the pulsar has higher sensitivity to high-frequency (periods less than $\lesssim$ a few yr) continuous wave gravitational-wave sources.
\item It is likely that many millisecond pulsars will exhibit timing noise over sufficiently long data spans \citep{Shannon10}.  Pulsars whose residuals are currently dominated by timing noise allow us to study this potentially sensitivity-limiting noise process.  A similar analysis on a large sample of normal pulsars \citep{lhk+10}  indicated that timing noise can be described as a two-state process with correlated pulse shape changes.  If such an effect (or something similar) can be identified for millisecond pulsar timing noise then such noise may potentially be mitigated \citep[cf.,][]{Oslowski11}.
\item For many of the DR2 pulsars, DM variations produce the largest timing residual deviations.  Most of these can be modelled, and hence removed, using multi-frequency observations. However, remnant variations will remain.  A study of ESE events, flux density variations and DM variations in a large sample of millisecond pulsars will enable optimal mitigation strategies to be developed and tested.

\end{itemize}

During most of the time period spanned by our DR2 observations, the Parkes telescope was the only southern-hemisphere telescope carrying out high-precision pulsar timing observations.  The observations of PSRs~J0437$-$4715, J0711$-$6830, J1017$-$7156, J1045$-$4509, J1125$-$6014, J1446$-$4701, J1545$-$4550, J1603$-$7202, J1732$-$5049, J2124$-$3358, J2129$-$5721 and J2241$-$5236 are therefore unique to this data set.

Since 2019 February (MJD~58526) the MeerTime project on MeerKAT \citep{bbb18} has been observing Southern pulsars. MeerKAT has a sensitivity five times that of Parkes in the 20\,cm observing band.  A total of 5000\,hr over 5\,yr has been allocated to the primary pulsar timing projects, but this time must be distributed between studies of millisecond pulsars, normal pulsars, globular pulsar studies and relativistic binaries.  MeerKAT PTA observations will have 300 to 400\,hr per year  (compared with almost 1000\,hr for Parkes).   For observational efficiency, pulsars that are ``jitter dominated'' with MeerKAT  (and Parkes) should continue to be observed with Parkes \citep{Shannon14}.  
With the UWL receiving system, Parkes now  has  a wider bandwidth (704 to 4032\,MHz, compared with 856 to 1712\,MHz for current observations with MeerKAT), and established polarimetric calibration methods. We also note also that MeerKAT and Parkes are separated by 130$^\circ$ in longitude,  giving extended (or even continuous) time coverage for pulsar observations.

\section{Summary and Conclusions}

We have presented a new data release (DR2) from the PPTA project, with observations spanning 2004 to 2018.  
The data was produced using a new processing pipeline intended for high-precision pulsar timing observations. 

We expect that DR2 will be used for numerous scientific applications including the hunt for nanohertz gravitational waves, the study the size and motions of planets in our solar system, the search for irregularities in terrestrial time standards and to obtain a detailed understanding of the individual millisecond pulsars.  We have already started work on detailed modelling of the white and low-frequency timing noise in the data, on bounding gravitational wave signals and on improving millisecond pulsar timing model parameters.  These results will be presented in future papers.

The data release is stand-alone and publicly available for download and use (\url{https://doi.org/10.25919/5db90a8bdeb59}).  We anticipate that it will be combined with the International Pulsar Timing Array (IPTA) data sets and hence become part of the most sensitive data sets available for PTA research.

Our pipeline is based around input data files that can be loaded and processed using the \textsc{psrchive} software suite and hence the pipeline can be used for many of the pulsar projects being carried out at Parkes \cite[including P574 and P789;][]{Johnston18,psj+19}. It would also work with most data sets from the IPTA telescopes as well as the emerging telescope systems such as FAST and MeerKAT, and the planned Square Kilometre Array, provided these are based on \textsc{psrchive}-compatible file structures. These high-gain telescopes are likely to carry out relatively short observations of thousands of pulsars (and hundreds of millisecond pulsars).  In contrast to the current PPTA sample of pulsars, such data sets cannot be curated carefully by hand and automated pipelines will be necessary.  As these telescopes will be the most sensitive available, but also will not have significant overlap in sky coverage, it will become challenging to confirm any results from these new telescopes using a different system.  It is therefore important that the processing is carried out in a transparent manner that provides reproducibility and provenance. The pipeline procedures developed for DR2 are a step in this direction.

It would be possible to make incremental improvements to the data set. For example, by combining with IPTA partner data sets, it will likely be possible to determine the fitted jumps to higher precision. Similarly, further investigations with  RFI excision methods \citep{Lazarus16} and wide-band timing methodologies \citep{Pennucci14} are likely to be beneficial.
Significant improvements in the quality or extent of the PPTA DR2 data set will be the subject of future data releases.

 We have now transitioned to the UWL receiver system.  Instead of carrying out observations first in the 20-cm band and then again with the dual-band receiver we can now obtain the entire band from 704\,MHz to 4032\,MHz) in a single observation with higher sensitivity, especially at the high end of the band.  This enables us to at least halve the observing time required to achieve similar results to those presented in this paper as well as ensuring that we are less affected by diffractive scintillation. We have therefore been able to increase the number of pulsars being observed by the PPTA. Even with the significantly improved instrumentation now available at the Parkes observatory, the legacy data sets presented here have enormous value.  Most PTA-style projects rely primarily on high-precision timing over long data spans and therefore these DR2 observations will continue to analysed as part of future pulsar timing programs.  We have an approximately 1-year overlap between the old and new systems that will enable the new data to be joined to DR2 with precisely measured timing offsets.

\section*{Acknowledgments}

We thank the referee, J. Verbiest, for suggestions and comments that improved the manuscript. 
The Parkes radio telescope is part of the Australia Telescope, which is funded by the Commonwealth Government for operation as a National Facility managed by CSIRO. This paper includes archived data obtained through the Australia Telescope Online Archive and the CSIRO Data Access Portal (http://data.csiro.au).  Work at NRL is supported by NASA.  M.B., S.O., and R.M.S., acknowledge Australian Research Council grant FL150100148.  R.M.S. also acknowledges funding support through Australian Research Council Future Fellowship FT190100155.    P.D.L. is supported by Australian Research Council Future Fellowship FT160100112 and Discovery Project DP180103155. Parts of this research were conducted by the Australian Research Council Centre of Excellence for Gravitational Wave Discovery (OzGrav), through project number CE170100004.
H.G.W. is supported by the National Natural Science Foundation of China (11573008).
J.W. is supported by the Youth Innovation Promotion Association of Chinese Academy of Sciences.   This research has made use of NASA's Astrophysics Data System.

\appendix
\onecolumn

\section{Example of a psrsh script}\label{sec:psrsh}

We provide a typical \textit{psrsh} script that is available in our data collection and corresponds to processing an observation of PSR~J1603$-$7202 (the observation file name is t081012\_075319.rf).  We note that this script has specific directory structures for our computer systems, but the use of a simple Python module specifying such structure means this can easily be modified for other systems.  (We also note that here we have updated the details of our file structure to a simpler system; the \textit{psrsh} scripts in the data collection contain the full paths).  A \verb|#| symbol indicates a comment line. These are automatically generated by the pipeline scripts. The \textit{psrsh} script is generated by a Python framework and associated metadata, and is then run automatically by the PPTA DR2 processing pipeline.

\begin{verbatim}
load /rawData/J1603-7202/t081012_075319.rf
# * Set 1050CM feed parameters.
edit rcvr:basis=lin
edit rcvr:hand=-1
edit rcvr:sa=0.0
edit rcvr:rph=0.0
# * Replace aliased pulsar name with the preferred one.
# * pdfb4.*_1024_[1,2]... chan 2.230  0.188   54751.30 55319.18  081012 100503
# * PDFB4_2bin issue 54751.30 to 55319.18
edit ext:stt_offs=0.000965354880627
# * Apply pdfb4_reset_bug flag to PDFB4 data between MJD 54731 and 54751.30.
zap edge 0.05
cal load /calData/J1603-7202/t081012_075036.cf.dzT
cal
cal frontend
cal load /fluxcalData/pdfb4_10cm_1.avfluxcal
cal flux
edit dm=38.0488
install par /ephemerisData/J1603-7202.par
# Compute full resolution dynamic spectrum.
dynspec /output/t081012_075319.rf.pcm.dynspec
   /template/J1603-7202_10cm_ana_PDFB2.std
tscrunch 8
fscrunch 32
unload /output/J1603-7202/t081012_075319.rf.pcm.dzt8f32

tscrunch 1
fscrunch 1
unload /output/J1603-7202/t081012_075319.rf.pcm.dzTF

pscrunch
unload /output/J1603-7202/t081012_075319.rf.pcm.dzTFp
\end{verbatim}

The overall organization follows a reduction from the original, high-frequency uncalibrated data to final, highly-averaged profiles.  There are multiple outputs, and care is taken to avoid redundant processing.  In brief, this script loads the original data file (the \verb|load| command) and ensures that the feed parameters are set (\verb|edit rcvr:*|) so that calibrated data are consistent with
the PSR/IEEE convention \citep{vmjr10}. Any known issues in the time stored in the header are fixed (\verb|edit ext:stt_offs|) and then 5\% of the band edges are removed (\verb|zap|).  The calibration file is loaded (both the pulsed calibrator prior to the observation and the flux calibration information) and the data file is calibrated (the \verb|cal| set of commands; note that calling \verb|cal| without any arguments applies the loaded polarization calibrator).  The dispersion measure is set  using \verb|edit dm| and an initial parameter file used to ensure that all files are processed using the same set of parameters for a given pulsar (\verb|install par|).  A dynamic spectrum of the calibrated file is then produced (\verb|dynspec|), and the output profile averaged to 8 time sub-integrations and 32 frequency channels (\verb|tscrunch|, \verb|fscrunch|).  An output file with file extension \verb|dzt8f32| is produced and then the profile is further averaged in time and frequency (to produce a file with extension \verb|dzTF|) and then finally a Stokes I profile is output (with extension \verb|dzTFp|).

\section{Example of a rule set used by the pipeline}\label{sec:ruleSet}

The following is an example of a set of rules related to manual RFI removal. Some manual RFI removal was necessary.  By following these rule sets it should be possible to reproduce our RFI flagging.
\verb|-z| indicates the removal of a particular channel while \verb|-w| a particular sub-integration.  The comments written (lines preceded by \verb|#|) are generated by the pipeline. 

\begin{verbatim}
# Some bands are relatively clean and we don't want to run automatic
# RFI zapping.  So just enumerate affected subints here.

# Format:
# basename_of_observation 1,2,30
@manual

# PSR~J0437-4715
w040504_051040.rf -w 12-15
w040815_200745.rf -z 146
w040816_182842.rf -z 146
w040816_190614.rf -z 146
w040816_202134.rf -z 146
w040816_205514.rf -z 146
w040830_184220.rf -z 146
w040831_172241.rf -z 146 -w 19,20
w040831_185715.rf -z 146
w040901_173923.rf -z 146
w040901_185956.rf -z 146
w041023_123208.rf -w 32
w050510_070547.rf -w 6

m2004-06-03-02:56:00.rf 19-24
n2004-08-28-20:19:59.rf -z 0-11 -w 0,3,4,5,14
...
\end{verbatim}

The following is an example indicating which data files are affected by catastrophic RFI and cannot be recovered.  The rule \verb|@catastrophic_rfi| maps to a Python function responsible for treating the tabulated observations:

\begin{verbatim}
# Observations so strongly affected by RFI that every subint/channel
# is affected.  E.g., early occurrences of aircraft-radar interference.

# These data are a total write-off and should not be processed.

# The directive below applies flag to data.

# NB that these pulsars are specifically those used in PPTA.  It's
# sensible to separate them since there are so many observations per pulsar.

@catastrophic_rfi

# PSR~J0613-0200
p150831_225744.rf
p150831_230240.rf

# PSR~J0711-6830
p151221_113232.rf
...
\end{verbatim}

\section{Example of jumps in parameter files}\label{sec:jumps}

The parameter files in our data release are not suitable for detailed studies of the pulsars themselves. Instead they provide an initial set of parameters along with a set of ``jumps'' that correspond to a given observing system, or time range.  As an example, we show the relevant ``jump'' parameters for PSR~J0711$-$6830:

\begin{verbatim}
JUMP -h 20CM_H-OH_CPSR2m -5.0824507989795e-07 1
JUMP -h 20CM_H-OH_CPSR2n 4.1751150972045e-07 1
JUMP -h 20CM_H-OH_PDFB4 5.8676673063388e-07 1
JUMP -h 20CM_H-OH_PDFB1 -1.5895467008059e-06 1
JUMP -h 20CM_MULTI_CPSR2m -1.2285860948976e-06 1
JUMP -h 20CM_MULTI_CPSR2n -1.2852059355821e-06 1
JUMP -g 50CM_CPSR2 1.9184474567463e-06 1
JUMP -g 10CM_WBCORR -1.2985508878209e-07 1
JUMP -h 20CM_MULTI_PDFB1 -1.608849753432e-06 1
JUMP -g 10CM_PDFB1 -6.9361254364538e-07 1
JUMP -g 20CM_PDFB2 -9.0880524215454e-07 1
JUMP -g 10CM_PDFB2 -2.3135497068471e-06 1
JUMP -g 20CM_PDFB3 -1.4283590712073e-06 1
JUMP -g 40CM_PDFB3 -3.6219962470862e-07 1
JUMP -g 40CM_CASPSR -2.5577420710035e-06 1
JUMP -g 20CM_PDFB4 -1.9669236905371e-06 1
JUMP -j 10CM_55319_PDFB4 8.4851151925669e-07 1
JUMP -j 20CM_55319_PDFB4 6.0052124013235e-07 1
JUMP -j 20CM_55319_PDFB3 5.5713865081463e-08 1
JUMP -j 40CM_55319_PDFB3 -2.2188193035067e-06 1
JUMP -cpsr2_50cm 1 -2.36e-06 0
JUMP -cpsr2m_1341 1 -1.72e-06 0
JUMP -cpsr2n_1405 1 -1.8e-06 0
JUMP -dfb3_J0437_55319_56160 1 2.2e-07 0
JUMP -dfb3_J0437_56160_60000 1 4.5e-07 0
JUMP -pdfb1_512_ch 1 -1.22813e-05 0
JUMP -pdfb1_post_2006 1 -1.3e-07 0
JUMP -pdfb1_pre_2006 1 -1.13e-06 0
JUMP -pdfb2_1024_MHz 1 -5.435e-06 0
JUMP -pdfb2_256MHz_1024_ch 1 -1.1395e-05 0
JUMP -pdfb3_1024_MHz 1 1.03e-06 0
JUMP -pdfb3_256MHz_1024ch 1 4.295e-06 0
JUMP -pdfb3_256MHz_2048ch 1 8.32e-06 0
JUMP -pdfb3_64MHz_1024ch 1 1.494e-05 0
JUMP -pdfb3_64MHz_512ch 1 8.9e-06 0
JUMP -pdfb4.*_1024_[1,2]... 1 2.23e-06 0
JUMP -pdfb4_256MHz_1024ch 1 5.05e-06 0
JUMP -pdfb4_55319_56055_cals 1 9.27e-07 0
JUMP -pdfb4_56055_56110_cals 1 3.82e-07 0
JUMP -pdfb4_56110_56160_cals 1 5.41e-07 0
JUMP -wbb_c_config 1 3.8e-07 0
JUMP -caspsr_55500_57575_cals 1 -1.31e-06 0
JUMP -pdfb4_57575_65000_cals 1 1.71e-07 0
JUMP -pdfb4_56160_57575_cals 1 4.25e-07 0
JUMP -caspsr_57575_65000_cals 1 -1.533e-06 0
\end{verbatim}

The final column (a zero or a one) indicates whether the jump has been measured precisely and should be held fixed (a zero) or whether the jump should be fitted for as part of the timing procedure (a one). A complete list of offsets that have been measured and therefore should be held as fixed values in the timing model fitting process are listed in Table~\ref{tb:Jumps}.  Offsets that are allowed to vary in the timing model fit are  also listed in Table~\ref{tb:Jumps}. Note that the  values for these jumps are often significantly different between the pulsars.

The jump naming convention reflects the a changes in the type of instrumental offsets experienced over the projects duration. 
Prior to MJD~55318 the digital filterbanks  (PDFB1,PDFB2,PDFB3, PDFB4) and wide-band correlator had instrumental jumps that depended on the specific frequency, and phase bin resolution configuration.  This was removed with firmware updates to the system.  Post MJD~55318 timing offsets are related to physical movements of systems or changes in cable lengths in signal paths.  We refer the reader to M+13 for discussion of the origin of instrumental offsets.

\begin{footnotesize}
\begin{table}\caption{Timing offsets in the PPTA DR2}\label{tb:Jumps}
\begin{tabular}{ll|cl}
\hline
\multicolumn{2}{c}{Fixed Offsets} & \multicolumn{2}{c}{Fitted Offsets} \\
Flag ID & Offset ($\mu$s) & Flag ID & Flag value\\
\hline
\verb|-caspsr_55500_57575_cals| & $-$1.31	 & -g & \verb|10CM_CPSR2| \\
\verb|-caspsr_57575_65000_cals| & $-$1.533	 & -g & \verb|10CM_PDFB1| \\
\verb|-cpsr2_50cm| & $-$2.36	 & -g & \verb|10CM_PDFB2| \\
\verb|-cpsr2m_1341| & $-$1.72	 & -g & \verb|10CM_WBCORR| \\
\verb|-cpsr2n_1405| & $-$1.8	 & -g & \verb|20CM_PDFB2| \\
\verb|-dfb3_J0437_55319_56160| & 0.22	 & -g & \verb|20CM_PDFB3| \\
\verb|-dfb3_J0437_56160_60000| & 0.45	 & -g & \verb|20CM_PDFB4| \\
\verb|-pdfb1_128_ch| & $-$3.5547	 & -g & \verb|40CM_CASPSR| \\
\verb|-pdfb1_2048_ch| & $-$46.8829	 & -g & \verb|40CM_PDFB3| \\
\verb|-pdfb1_32_ch| & $-$1.2969	 & -g & \verb|50CM_CPSR2| \\
\verb|-pdfb1_512_ch| & $-$12.2813	 & -h & \verb|20CM_H-OH_CPSR2m| \\
\verb|-pdfb1_j1909_2006| & 12.2813	 & -h & \verb|20CM_H-OH_CPSR2n| \\
\verb|-pdfb1_post_2006| & $-$0.13	 & -h & \verb|20CM_H-OH_PDFB1| \\
\verb|-pdfb1_pre_2006| & $-$1.13	 & -h & \verb|20CM_H-OH_PDFB4| \\
\verb|-pdfb2_1024_MHz| & $-$5.435	 & -h & \verb|20CM_MULTI_CPSR2m| \\
\verb|-pdfb2_256MHz_1024_ch| & $-$11.395	 & -h & \verb|20CM_MULTI_CPSR2n| \\
\verb|-pdfb2_256MHz_2048_ch| & $-$14.35	 & -h & \verb|20CM_MULTI_PDFB1| \\
\verb|-pdfb2_256MHz_512ch| & $-$4.75	 & -j & \verb|10CM_55319_PDFB4| \\
\verb|-pdfb3_1024_256_512| & 2.45	 & -j & \verb|10CM_PDFB1_JUMP| \\
\verb|-pdfb3_1024_MHz| & 1.03	 & -j & \verb|20CM_55319_PDFB3| \\
\verb|-pdfb3_256MHz_1024ch| & 4.295	 & -j & \verb|20CM_55319_PDFB4| \\
\verb|-pdfb3_256MHz_2048ch| & 8.32	 & -j & \verb|20CM_PDFB1_1433_JUMP| \\
\verb|-pdfb3_64MHz_1024ch| & 14.94	 & -j & \verb|20CM_PDFB1_JUMP| \\
\verb|-pdfb3_64MHz_512ch| & 8.9	 & -j & \verb|40CM_55319_PDFB3| \\
\verb|-pdfb4.*_1024_[1,2]...| & 2.23 \\
\verb|-pdfb4_1024_256_512| & 3.23 \\
\verb|-pdfb4_2048_1024_1024| & 2.14 \\
\verb|-pdfb4_256MHz_1024ch| & 5.05 \\
\verb|-pdfb4_256MHz_2048ch| & 9.22 \\
\verb|-pdfb4_55319_56055_cals| & 0.927 \\
\verb|-pdfb4_56055_56110_cals| & 0.382 \\
\verb|-pdfb4_56110_56160_cals| & 0.541 \\
\verb|-pdfb4_56160_57575_cals| & 0.425 \\
\verb|-pdfb4_57575_65000_cals| & 0.171 \\
\verb|-wbb256_1024_128_3p| & $-$0.62 \\
\verb|-wbb256_512_128_3p_b| & $-$0.62 \\
\verb|-wbb_c_config| & 0.38 \\
\hline
\end{tabular}
\end{table}
\end{footnotesize}

\section{Example of pulse ToAs}\label{sec:exampleToA}

The following is part of PSR~J0711$-$6830.tim, the arrival time file corresponding to observations of PSR~J0711$-$6830.

\begin{verbatim}
t180424_104338.rf.pcm.dzTf8p 2908.50000000 58232.46932869677628375 2.83300 pks 
 -f 1050CM_PDFB4 -g 10CM_PDFB4 -h 10CM_1050CM_PDFB4 -fe 1050CM -be PDFB4 
 -B 10CM -length 3839.99 -tobs 3839.99 -bw 1024.00 -nchan 1024 -pta PPTA 
 -tmplt J0711-6830_10cm_ana_PDFB2.std -pdfb4_57575_65000_cals 1 -projid P456
 -beconfig pdfb4_1024_1024_1024 -snr 63.15 -gof 0.93 -group PDFB_10CM 
 -v 10CM_PDFB4 -j 10CM_55319_PDFB4 
t180424_104338.rf.pcm.dzTf8p 3036.50000000 58232.46932868819421358 3.30300 pks 
 -f 1050CM_PDFB4 -g 10CM_PDFB4 -h 10CM_1050CM_PDFB4 -fe 1050CM -be PDFB4
 -B 10CM -length 3839.99 -tobs 3839.99 -bw 1024.00 -nchan 1024 -pta PPTA 
 -tmplt J0711-6830_10cm_ana_PDFB2.std -pdfb4_57575_65000_cals 1 -projid P456
 -beconfig pdfb4_1024_1024_1024 -snr 51.96 -gof 0.96 -group PDFB_10CM 
 -v 10CM_PDFB4 -j 10CM_55319_PDFB4 
t180424_104338.rf.pcm.dzTf8p 3164.50000000 58232.46932868052305921 3.69500 pks 
 -f 1050CM_PDFB4 -g 10CM_PDFB4 -h 10CM_1050CM_PDFB4 -fe 1050CM -be PDFB4 
 -B 10CM -length 3839.99 -tobs 3839.99 -bw 1024.00 -nchan 1024 -pta PPTA 
 -tmplt J0711-6830_10cm_ana_PDFB2.std -pdfb4_57575_65000_cals 1 -projid P456
 -beconfig pdfb4_1024_1024_1024 -snr 49.66 -gof 0.97 -group PDFB_10CM 
 -v 10CM_PDFB4 -j 10CM_55319_PDFB4 
t180424_104338.rf.pcm.dzTf8p 3292.50000000 58232.46932867376629517 3.43400 pks 
 -f 1050CM_PDFB4 -g 10CM_PDFB4 -h 10CM_1050CM_PDFB4 -fe 1050CM -be PDFB4 
 -B 10CM -length 3839.99 -tobs 3839.99 -bw 1024.00 -nchan 1024 -pta PPTA 
 -tmplt J0711-6830_10cm_ana_PDFB2.std -pdfb4_57575_65000_cals 1 -projid P456
 -beconfig pdfb4_1024_1024_1024 -snr 51.84 -gof 1.03 -group PDFB_10CM 
 -v 10CM_PDFB4 -j 10CM_55319_PDFB4 
\end{verbatim}

The file is in TEMPO2 format and details for each observation are given on a single line.  The first five columns are required by TEMPO2.  These are (1) the filename, (2) the observing frequency corresponding to the ToA determination (MHz), (3) the arrival time (MJD), (4) the ToA uncertainty ($\mu s$) and (5) the site observing code (in all cases this is ``pks'' corresponding to the Parkes telescope).

The remaining columns contain TEMPO2 flags and their values. These are:
\begin{itemize}
    \item -f: a flag containing both the front end receiver (e.g., 1050CM) and the backend instrument (e.g., PDFB4).
    \item -g: similar to ``-f", but specifying the observing band (e.g., 10CM) instead of the receiver used.
    \item -h: a flag concatenating the receiver, band and backend instrument.
    \item -fe: the name of the front-end receiver system
    \item -be: the name of the back-end instrument
    \item -B: the observing band
    \item -length: the total observation length (seconds)
    \item -tobs: 
    \item -bw: the bandwidth of the observation (MHz)
    \item -nchan: the number of frequency channels in the original observation.
    \item -pta: the pulsar timing array recording the data (in our case all these are set to ``PPTA'')
    \item -tmplt: the name of the template profile used when forming the arrival times.
    \item -projid: the Parkes observing project (typically P456).
    \item -beconfig: the backend configuration, which is typically the \verb|<instrument>_<nbin>_<bw>_<nchan>|
    \item -snr: the S/N of the resulting pulse profile
    \item -gof: the goodness-of-fit during the template matching procedure 
    \item -group: a grouping of specific frontend and backend instruments
    \item -v: \verb|<band>_<instrument>|
    \item -j: fitted jump
\end{itemize}
Some observations also contain specific flags (such as \verb|-pdfb4_57575_65000_cals| in the example above) that are used to identify specific groups of data points for flagging or for applying time offsets.  Full details of the mapping between delay flags and signal processors, configurations and dates, are available from the ``ruleSets'' directory in our public data collection.

\section{Example and use of a tempo2 select file}\label{sec:selectFiles}

TEMPO2 selection file (``select'' files) are used to enable observations identified via specific instruments or flags to be removed and not included in subsequent analysis.

\begin{verbatim}
LOGIC -bad_config = 1 REJECT
LOGIC -caspsr_commissioning = 1 REJECT
LOGIC -corr_prob = 1 REJECT
LOGIC -dfb1_2007_offset = 1 REJECT
LOGIC -nocal = 1 REJECT
LOGIC -notiming = 1 REJECT
LOGIC -pdfb2_commissioning = 1 REJECT
LOGIC -pdfb3_commissioning = 1 REJECT
LOGIC -pdfb3_phase_lock = 1 REJECT
LOGIC -pdfb4_commissioning = 1 REJECT
LOGIC -pdfb4_phase_lock = 1 REJECT
LOGIC -pdfb4_reset_bug = 1 REJECT
LOGIC -pks_clk_prob = 1 REJECT
LOGIC -poor_profile = 1 REJECT
LOGIC -unusual_config = 1 REJECT
LOGIC -varying_profile = 1 REJECT
LOGIC -wbc_phase_lock = 1 REJECT
PROCESS freqpass 600 800
PROCESS freqpass 1200 1500
PROCESS freqpass 2800 3200
LOGIC -length < 300 REJECT
LOGIC -snr < 15 REJECT
PROCESS -fe DRAO REJECT
PROCESS -fe MARS REJECT
PROCESS -fe 13MM REJECT
PROCESS -fe GALILEO REJECT
\end{verbatim}

The first column contains a command.  The \verb|LOGIC| command parses a logical expression to decide on the output.  The \verb|PROCESS| command will carry out a particular processing task given some parameters.   For instance:

\begin{verbatim}
    LOGIC -unusual_config = 1 REJECT
\end{verbatim}
will reject (i.e., not include in further processing) all observations that have been identified with the \verb|-unusual_config| flag. 

\begin{verbatim}
    LOGIC -snr < 15 REJECT
\end{verbatim}
implies that any observation that has a \verb|-snr| flag that is smaller than 15 will be rejected.  The 
\begin{verbatim}
    PROCESS freqpass 600 800
\end{verbatim}
command will ``pass'' (i.e., not reject) any observations in which the observing frequency is within the range of 600 to 800\,MHz. The selection file can be implemented using a \textsc{tempo2} command line that includes the \verb|-select <filename>| command line argument.

\section{Author contributions and affiliations}

The PPTA project has a large number of team members. The authors (in the initial non-alphabetical section) led specific aspects of the work to develop the data release presented here, or carried out a large fraction of the observations.  The remaining author list is in alphabetical order and includes individuals who were involved in the  project during the time spanned by the observations reported here.

The author affilliations are as follows:

\noindent
{$^{1}$ Space Science Division, Naval Research Laboratory, Washington, DC 20375-5352, USA}  \\
{$^{2}$ Centre for Astrophysics and Supercomputing, Swinburne University of Technology, P.O. Box 218, Hawthorn, Victoria 3122, Australia}  \\
{$^{3}$ Australian Research Council Centre of Excellence for Gravitational Wave Discovery (OzGrav)} \\
{$^{4}$ CSIRO Astronomy and Space Science, Australia Telescope National Facility, PO~Box~76, Epping NSW~1710, Australia}  \\
{$^{5}$ CSIRO Scientific Computing, Australian Technology Park, Locked Bag 9013, Alexandria, NSW 1435, Australia} \\
{$^{6}$ University of Chinese Academy of Sciences, Beijing 100049, China} \\
{$^{7}$ Purple Mountain Observatory, Chinese Academy of Sciences, Nanjing 210008, China} \\
{$^{8}$ International Centre for Radio Astronomy Research, University of Western Australia, Crawley, WA 6009, Australia}\\
{$^{9}$ Institute for Radio Astronomy \& Space Research, Auckland University of Technology, Private Bag 92006, Auckland 1142, New Zealand} \\
{$^{10}$ International Centre for Radio Astronomy Research, Curtin University, Bentley, Western Australia 6102, Australia}\\
{$^{11}$ Department of Electrical and Computer Engineering, University of California at San Diego, La Jolla, CA 92093, USA}\\
{$^{12}$ CSIRO Information Management \& Technology, GPO Box 1700, Canberra ACT 2601} \\
{$^{13}$ School of Physics and Astronomy, Monash University, VIC 3800, Australia} \\
{$^{14}$ CSIRO Astronomy and Space Science, Parkes Observatory, 473 Telescope road, Parkes NSW, 2870}\\
{$^{15}$ Jodrell Bank Centre for Astrophysics, Department of Physics and Astronomy, The University of Manchester, Alan Turing Building, Oxford Road, Manchester, M13 9PL, UK
} \\
{$^{16}$ School of Physics and Electronic Engineering, Guangzhou University, Guangzhou 510006, China}\\
{$^{17}$ Xinjiang Astronomical Observatory, Chinese Academy of Sciences, 150 Science 1-Street, Urumqi, Xinjiang 830011, China} \\
{$^{18}$ National Astronomical Observatories, Chinese Academy of Sciences, A20 Datun Road, Chaoyang District, Beijing 100101, China}
\label{lastpage}

\end{document}